\documentclass[3p,preprint]{elsarticle}

\usepackage{color}
\usepackage[utf8]{inputenc}
\usepackage{graphicx}
\usepackage{amsmath}
\usepackage{amssymb,bm, physics, dsfont}
\usepackage{calligra}
\usepackage{listings}
\usepackage{longtable}
\usepackage{footnote}
\makesavenoteenv{tabular}
\makesavenoteenv{table}
\usepackage{threeparttable}
\usepackage{here}
\usepackage{ulem}

\usepackage{hyperref}
\hypersetup{
  pdfnewwindow=true, colorlinks=true,
  linkcolor=blue, anchorcolor=blue,
  citecolor=blue, filecolor=blue,
  menucolor=blue, urlcolor=blue}

\graphicspath{{./Fig/}}

\newcommand{\mZ}{\mathbb{Z}}
\newcommand{\bx}{\bm{x}}
\newcommand{\bk}{\bm{k}}

\newcommand{\bt}{\bm{t}}

\newcommand{\bR}{\bm{R}}
\newcommand{\bG}{\bm{G}}

\newcommand{\calG}{\mathcal{G}}
\newcommand{\calGk}{\mathcal{G}_{\bm{k}}}

\newcommand{\vk}{\bm{k}}

\newcommand{\qeirreps}{\texttt{qeirreps}}

\begin{document}

\begin{frontmatter}

\author[UTokyo]{Akishi Matsugatani\corref{cor1}}\ead{matsugatani@cmt.t.u-tokyo.ac.jp}
\author[UTokyo]{Seishiro Ono}
\author[RIKEN]{Yusuke Nomura}
\author[UTokyo]{Haruki Watanabe}
\address[UTokyo]{Department of Applied Physics, University of Tokyo, Tokyo 113-8656, Japan}
\address[RIKEN]{RIKEN Center for Emergent Matter Science, 2-1 Hirosawa, Wako, Saitama 351-0198, Japan}

\date{}
\title{\texttt{qeirreps}: an open-source program for \textsc{Quantum ESPRESSO} to compute irreducible representations of Bloch wavefunctions}

\begin{abstract}
Bloch wavefunctions in solids form a representation of crystalline symmetries. Recent studies revealed that symmetry representations in band structure can be used to diagnose the topological properties of weakly interacting materials. In this work, we introduce an open-source program \texttt{qeirreps} that computes the representation characters in a band structure based on the output file of \textsc{Quantum ESPRESSO}. Our program also calculates the  $\mZ_4$ index, i.e., the sum of inversion parities at all time-reversal invariant momenta, for materials with inversion symmetry. When combined with the symmetry indicator method, this program can be used to explore new topological materials.
\end{abstract}


\end{frontmatter}

\noindent\textbf{PROGRAM SUMMARY}

\noindent\textit{Program title:} qeirreps

\noindent\textit{Catalogue identifier:} 



\noindent\textit{Licensing provisions:} GNU General Public Licence 3.0




\noindent\textit{Programming language:} Fortran 90

\noindent\textit{Computer:} any architecture with a Fortran 90 compiler

\noindent\textit{Operating system:} Unix, Linux


\noindent\textit{RAM:} Variable, depends on the complexity of the problem


\noindent\textit{External routines/libraries used:} 
\begin{itemize}
\item BLAS (http://www/netlib.org/blas)
\item LAPACK (http://www.netlib.org/lapack)
\end{itemize}

\noindent\textit{Nature of problem:} Irreducible representations of Bloch wavefunctions

\noindent\textit{Solution method:}  Linear algebra calculation for Bloch wavefunctions 

\noindent\textit{Running time:} 1 min - 1 h (strongly depends on the complexity of the problem)

\section{Introduction}
\label{sec1}
One of the main goals in condensed matter physics is to predict the properties of materials of our interest by solving the Schr\"odinger equation.  
In practice, this challenging problem is reduced to a manageable one in two ways.  The first is to map the interacting system to a free (i.e., noninteracting) electronic system, as is done in the Kohn-Sham density functional theory (DFT)~\cite{KS}.  The other simplification utilizes the symmetry of the system. For example, the lattice translation symmetry of an ideal crystal that is free from impurities or disorders allows us to block-diagonalize the Hamiltonian by switching to the momentum space. The original problem of interacting electrons is thereby transformed into the one within the standard band theory. The electronic structure of an enormous number of weakly interacting materials has been successfully computed in this way.

In addition to the lattice translation symmetry, a perfect crystal tends to have other symmetries such as a spatial inversion and a discrete rotation. The set of symmetry operations of a crystal forms a group, called space group. In three dimensions, there are 230 different space groups, and the spatial symmetry of the crystal at work falls into one of them.  Crystalline symmetries help us to a better understanding on the electronic band structure. Bloch wavefunctions form a representation of the space group, and the representation puts constraints on how many energy bands degenerate at each momentum and how energy bands at different momenta connect with each other.  Furthermore,  there has been an increasing number of evidence that the space group representation also informs us of the topological aspects of the Bloch wavefunction.

Since the discovery of the $\mZ_2$ topological insulator~\cite{PhysRevLett.95.146802,PhysRevLett.95.226801}, topological materials have attracted researchers around the world. 
In the early stage of the study of topological phases, the focus was on the phases protected by internal symmetries: the time-reversal, the particle-hole, and the chiral symmetry~\cite{Bernevig1757, Konig766,PhysRevB.74.195312,PhysRevLett.98.106803, PhysRevB.75.121306,Fu-Kane, Hsieh:2008aa,PhysRevLett.100.236601,PhysRevB.78.195424,Hsieh:2009aa,Xia:2009aa,PhysRevB.79.195322,Yu61,PhysRevLett.107.136603,Chang167}. These pioneering studies were united into the celebrated topological periodic table~\cite{PhysRevB.78.195125, doi:10.1063/1.3149495,Ryu_2010}.  They were followed by a large number of studies that take into account the various types of crystalline symmetries~\cite{PhysRevLett.106.106802,Freed2013,PhysRevB.88.125129, PhysRevB.88.075142,PhysRevB.90.205136, PhysRevB.90.165114,PhysRevB.91.155120,PhysRevB.93.195413,PhysRevB.95.235425, Wang:2016aa,PhysRevB.91.161105, PhysRevX.7.011020, PhysRevX.7.041069, PhysRevX.8.011040,PhysRevB.96.205106, Shiozaki2018, Song2018}.  It turned out that there exist various novel topological phases protected by crystalline symmetries~\cite{PhysRevLett.106.106802,PhysRevB.90.165114,PhysRevB.91.155120,PhysRevB.93.195413,PhysRevB.95.235425,Wang:2016aa,PhysRevLett.119.246402,PhysRevLett.119.246401,Benalcazar61,Fang2017}, such as mirror Chern insulators~\cite{Tanaka:2012aa,Hsieh:2012aa}, higher-order topological insulators~\cite{PhysRevLett.119.246402,Benalcazar61,Fang2017,Schindlereaat0346,PhysRevB.96.245115, PhysRevB.98.205129, PhysRevX.9.011012}, and topological semimetals of the Dirac~\cite{PhysRevLett.108.140405} and Weyl~\cite{PhysRevB.83.205101} type.  These topological materials host robust surface states and exhibits unique bulk responses, which could be leveraged for new low-power devices. Therefore, discovering new candidates of topological materials is one of the important tasks both for the fundamental physics and for the application.

At this moment, listing up the full set of topological invariants, for a given symmetry setting, that completely characterize all possible topological phases is still a pending problem.  Furthermore, even if we know the mathematical definition of topological invariants, it is sometimes difficult to compute them directly using their definitions~\cite{PhysRevB.91.155120, Wang:2016aa,PhysRevB.91.161105}.  These issues can be resolved with the help of crystalline symmetries.  By using the information of the representations of the Bloch wavefunctions at a set of crystalline momenta, it is sometimes possible to judge the topology encoded in the Bloch wavefunctions quite easily.  The prototypical example is the Fu-Kane formula~\cite{Fu-Kane}, which determines the $\mathbb{Z}_2$ invariant based on the inversion parities. Recently, this idea has been extended to a wider class of topological (crystalline) insulators and semimetals in all 230 space groups. The generalized schemes are nowadays known as the method of ``symmetry indicators''~\cite{Po2017,SI_Adrian} and ``topological quantum chemistry''~\cite{TQC}.  Their usefulness in the search for realistic topological materials was clearly demonstrated by the recent comprehensive survey of topological materials among existing databases of inorganic substances~\cite{Tang2019_NP,Tangeaau8725,Zhang2019,Vergniory2019,Tang2019}.  As a result of this survey, thousands of candidates of topological (crystalline) insulators and semimetals have been identified.  The theory of symmetry indicators has also been extended to magnetic space groups~\cite{Watanabeeaat8685} and superconductors~\cite{Ono-Watanabe2018,Ono-Yanase-Watanabe2019,Skurativska2019,1907.13632, Ono-Po-Watanabe2020, SI_Luka}, and further investigation of topological magnetic materials and superconductors is awaiting us.

Having in mind these applications, it is evidently important to make it possible to automatically compute irreducible representations of Bloch wavefunctions by using DFT. 
Although the authors of Refs.~\cite{Tang2019_NP,Tangeaau8725,Zhang2019,Vergniory2019,Tang2019, irvsp} implemented this function for WIEN2k~\cite{wien2k} and VASP~\cite{vasp}, these softwares require paid licenses.  In contrast, for the \textsc{Quantum ESPRESSO}~\cite{qe1,qe2}, which is a free, open-source package, the existing program can handle only one type of space groups, called ``symmorphic"  space groups, and does not work for the other type, called ``non-symmorphic." Given this situation, we develop an open-source code, named \texttt{qeirreps}, that works equally for symmorphic and non-symmorphic space groups.  This would allow everyone to compute the band structure and determine the irreducible representations by themselves, and when combined with the method of symmetry-indicator, would help our search of novel topological materials like topological superconductors as partially demonstrated in Ref.~\cite{Ono-Po-Shiozaki2020}.

The rest of this paper is organized as follows. 
In Sec.~\ref{sec2}, we briefly review the theoretical background of space groups and their representations. 
In Sec.~\ref{sec4}, we explain the installation and usage of \texttt{qeirreps}. 
After providing some examples: bismuth, silicon, NaCdAs, and PbPt$_3$ in Sec.~\ref{sec:ex}, we conclude in Sec.~\ref{sec5}.

\section{Theoretical background}
\label{sec2}
We start with reviewing the basics of space groups and their representations.  

\subsection{Space group action on the real and momentum space}
\label{subsec:SG}
Let us consider a space group $\calG$. An element $g \in \calG$ moves $\bx\in\mathbb{R}^3$ to
\begin{align}
\label{eq:SG}
g(\bx) &= p_g \bx + \bt_{g}\ \in \mathbb{R}^3,
\end{align}
where $p_g\in\text{O}(3)$ represents the point group part and $\bt_{g}\in\mathbb{R}^3$ represents the translation part of $g$. The product of two elements $g\in \calG$ and $g'\in \calG$ is defined by
\begin{align}
\label{product_G}
p_{gg'} = p_gp_{g'}, \quad \bt_{gg'} = p_g\bt_{g'} + \bt_{g}.
\end{align}
In general, $\bt_{g}$ is not necessarily a lattice vector. If we can choose the origin of $\bx$ in such a way that $\bt_{g}$ becomes a lattice vector simultaneously for all $\forall g \in \calG$,  the space group is symmorphic; otherwise, it is nonsymmorphic.  Nonsymmorphic space groups typically contain either screw axes or glide planes. 

A space group $\calG$ always has a subgroup  $T$ composed of lattice translations. An element $T_{\bR}$ of $T$ can be characterized by a lattice vector $\bR$.  
We introduce the wavevector  $\bk$ through the representation of the translation symmetry $U_{\bk}(T_{\bR}) = e^{-i \bk \cdot \bR}$.

An element $g\in \mathcal{G}$ changes $\bk$ to $g(\bk) = p_g \bk$. We say $\bk$ is invariant under $g$ when $g(\bk) = \bk + \bG$ with a reciprocal lattice vector $\bG$. 
For each $\bk$, the set of $g\in\mathcal{G}$ that leaves $\bm{k}$ invariant, i.e., $\left\{h \in \calG\ \vert\ h(\bk) = \bk + \bG\right\}$, forms a subgroup of $\calG$ called the little group $\mathcal{G}_{\bm{k}}$.
Since $\calGk$ is also an infinite group due to its translation subgroup $T$, sometimes the finite group $\calGk/T$, called ``little co-group~\cite{Bradley}," is discussed instead. In this work, we always consider the little group $\calGk$.

\subsection{Crystalline symmetries of band structures}
\label{subsec:BS}
The space group symmetry $\calG$ can be encoded in the single-particle Hamiltonian $H_{\vk}$ in momentum space by requiring
\begin{align}
\label{eq:symm_cond}
U_{\vk}(g)H_{\vk} &= H_{p_g\vk}U_{\vk}(g)
\end{align}
for each $g\in\calG$, where $U_{\vk}(g)$ is a unitary matrix forming a representation of $\mathcal{G}$. For spinful electrons, relevant representations become `projective' (in contrast to the ordinary `linear' representation) and $U_{\vk}(g)$'s satisfy
\begin{align}
\label{eq:rep_Gk}
U_{g'\vk}(g)U_{\vk}(g') &= \omega^{\text{sp}}(g,g')U_{\vk}(gg')
\end{align}
for $g,g'\in\calG$. 
The projective factor $\omega^{\text{sp}}(g,g')=\pm1$ is not unique, and we choose it as in Eq.~\eqref{spinprojective}.  For spin-rotation invariant systems, $\omega^{\text{sp}}(g,g')$ can be set $1$ by neglecting the spin degree of freedom.  

There is a way of treating projective representations of $\calG$ as linear representations of an enlarged group $\calG'$ who has the doubled number of elements. 
This method is known as `double group,' and is commonly used, for example, in Ref.~\cite{Bilbao}.  In this treatment, $\calG'$ contains both $\pm g$ for each element of $\calG$, and the product of $\eta g\in\calG'$  and $\eta' g'\in\calG'$ is defined by $\eta \eta'\omega^{\text{sp}}(g,g')gg'\in\calG'$. See \ref{appPR} for an example.
 Whether one handles projective representations directly, as we do in our program,  or via the double group method does not affect the physical conclusion.  See more detailed discussions on the relationship between projective representations and double groups in Refs.~\cite{PhysRevLett.117.096404,Watanabe201514665}.

The single-particle Hamiltonian $H_{\vk}$ can be diagonalized as
\begin{align}
\label{eq:Ham_TB_2}
H_{\vk} &= \Psi_{\vk}
\begin{pmatrix}
\epsilon_{\vk1}\mathds{1}_{1} & 0 & \cdots & 0 \\
0 &  \epsilon_{\vk2}\mathds{1}_{2}  & \cdots & 0 \\
\vdots & \vdots & \ddots & \cdots \\
0 & 0 & \cdots & \epsilon_{ \vk M_{\vk}}\mathds{1}_{M_{\vk}}
\end{pmatrix}\Psi_{\vk}^{\dagger},
\end{align}
where $\epsilon_{\vk n}$ ($n=1,2,\cdots,M_{\vk}$) is the $n$-th energy level of $H_{\vk}$ and $\mathds{1}_{n}$ is the identity matrix whose size is given by the order of degeneracy of $\epsilon_{\vk n}$.  The unitary matrix $\Psi_{\vk}$ is composed of all eigenvectors of $H_{\vk}$.  
When $h$ is an element of $\calGk$, $U_{\vk}(h)$ can also be block-diagonalized by $\Psi_{\vk}$ as
\begin{align}
\label{eq:Rep_TB}
U_{\vk}(h) &= \Psi_{\vk}
\begin{pmatrix}
U_{\vk1}(h) & 0 & \cdots & 0 \\
0 &  U_{\vk2}(h) & \cdots & 0 \\
\vdots & \vdots & \ddots & \cdots \\
0 & 0 & \cdots & U_{\vk M_{\vk}}(h) 
\end{pmatrix}\Psi_{\vk}^{\dagger} \quad \left(\forall h \in \calG_{\vk}\right).
\end{align}
Here, $U_{\vk n}(h)$ is the representation of $\calGk$ of the $n$-th band. Although the specific representation depends on the detailed choice of $\Psi_{\vk}$, its character 
\begin{align}
\chi_{\vk n}(h) &= \mathrm{tr}\left[U_{\vk n}(h) \right]
\end{align}
is basis independent.  The output of our program is the list of characters $\chi_{\vk n}(h)$ for each the energy level $\epsilon_{\vk n}$ at each high-symmetry point $\vk$.

In the absence of accidental degeneracy or symmetries other than $\mathcal{G}$, the representation $U_{\vk n}(h)$ is automatically irreducible. That is, the character $\chi_{\vk n}(h)$ of the $n$-th band coincides with one of the characters $\chi_{\vk}^\alpha(h)= \mathrm{tr}\left[U_{\vk}^\alpha(h) \right]$ of irreducible representations of $\calGk$.  Otherwise, $U_{\vk n}(h)$ can be decomposed into irreducible representations $U_{\vk}^\alpha$ as $U_{\vk n}=\oplus_\alpha n_{\vk n}^{\alpha}U_{\vk}^\alpha$, where the multiplicity of each irreducible representation is given by the formula
\begin{equation}
n_{\vk n}^{\alpha}=\frac{1}{|\calGk/T|}\sum_{h\in\calGk/T}\chi_{\vk}^\alpha(h){}^*\,\chi_{\vk n}(h).
\label{nknalpha}
\end{equation}
Our output file provides the list of $\chi_{\vk}^\alpha(h)$ and $n_{\vk n}^{\alpha}$ in \texttt{irreps\_list.dat} and \texttt{irreps\_number.dat}, respectively.~\footnote{Note that the output of our program is the character of representations of $\calGk$, not $\calGk/T$, and one should not confuse them. Irreducible representations of $\calGk$ and $\calGk/T$ are related to each other by a simple rule, and one can convert one to the other easily.}


\subsection{Application}
\label{subsec:SI}
The sum of integers $n_{\vk n}^{\alpha}$ in Eq.~\eqref{nknalpha} over all filled bands, i.e., $n_{\vk }^{\alpha}=\sum_{n:\text{filled}}n_{\vk n}^{\alpha}$, counts the irreducible representation $U_{\vk}^{\alpha}(h)$ below the Fermi level at each high-symmetry point $\vk$.  The integers $n_{\vk}^{\alpha}$  can be used, for example, to diagnose the topology of band insulators through the method of symmetry indicators~\cite{Po2017,SI_Adrian} or the topological quantum chemistry~\cite{TQC}. These methods diagnose the topology of the target material by comparing its irreducible representations with those of atomic insulators.
Useful formulas of topological indices in terms of $\{n_{\vk}^{\alpha}\}$ are provided in Refs.~\cite{PhysRevX.8.031069, PhysRevX.8.031070, QuantitativeMappings, Ono-Watanabe2018}.

For the user's convenience, we implemented the function that automatically computes the $\mZ_4$ index as the sum of the inversion parities of all occupied bands at all the time-reversal invariant momenta (TRIMs).
The $\mZ_4$ index can detect not only strong topological insulators but also topological crystalline insulators such as mirror Chern insulators or higher-order topological insulators~\cite{PhysRevX.8.031070, QuantitativeMappings}.  See Sec.~\ref{sub-sec:ex-bi} for an example.  
Other symmetry indicators can also be computed in the same way using the output of our program.  We discuss several examples in Sec.~\ref{sub-sec:ex-PbPt3}.

\subsection{Conventions}
\label{sec3}
Let us summarize our conventions that are necessary to compare the output of our program with that of others. 
There are three sources of ambiguities that affect the U(1) phase of the representation $U_{\vk}^{\alpha}(h)$: (i) the choice of representatives of $\mathcal{G}$, (ii) the choice of the coordinates, and (iii) the choice of spin rotation matrices. 
Readers not interested in the details can skip to Sec.~\ref{sec4}. 

\subsubsection{The choice of representatives of $\mathcal{G}$}
Although the number of elements of $\calG$ is infinite because of its translation subgroup, in the actual calculation, 
it is sufficient to discuss a finite number of elements by choosing one $g\in\calG$ for each $p_g$. 
This is because, if $g'$ differs from $g$ by its translation part (i.e., $p_{g'}=p_g$ and $\bm{t}_{g'}=\bm{t}_g+\bm{R}$ for a lattice vector $\bm{R}$), the little-group representation of $g'$ is simply given by $U_{\bm{k}}(g')=U_{\bm{k}}(g)e^{-i\bm{k}\cdot\bm{R}}$.
The choice of $g$'s is not unique, and in our program, they are automatically selected by \textsc{Quantum ESPRESSO} based on the input file.
The chosen elements of $\calG$ are stored in our output files: 
the list of $p_g$'s is in \texttt{pg.dat} and the list of the corresponding $\bm{t}_g$'s is in \texttt{tg.dat}.

\subsubsection{The choice of the coordinates}
When using \textsc{Quantum ESPRESSO}, one needs to prepare an input file that contains the information of the coordinates of atomic positions.
In \textsc{Quantum ESPRESSO}, the fixed point of point group symmetries is always set to $(0,0,0)$, and the input file must be carefully prepared.
For a given material, even after fixing the symmetry operation, there can still be multiple choices of the coordinates, and the irreducible representations can in general be affected by this choice.  
In our examples discussed in Sec.~\ref{sec:ex}, the information of the chosen coordinates is stored in the input file named \texttt{*.scf.in}.  

\begin{figure}[t]
	\begin{center}
		\includegraphics[width=0.75\columnwidth]{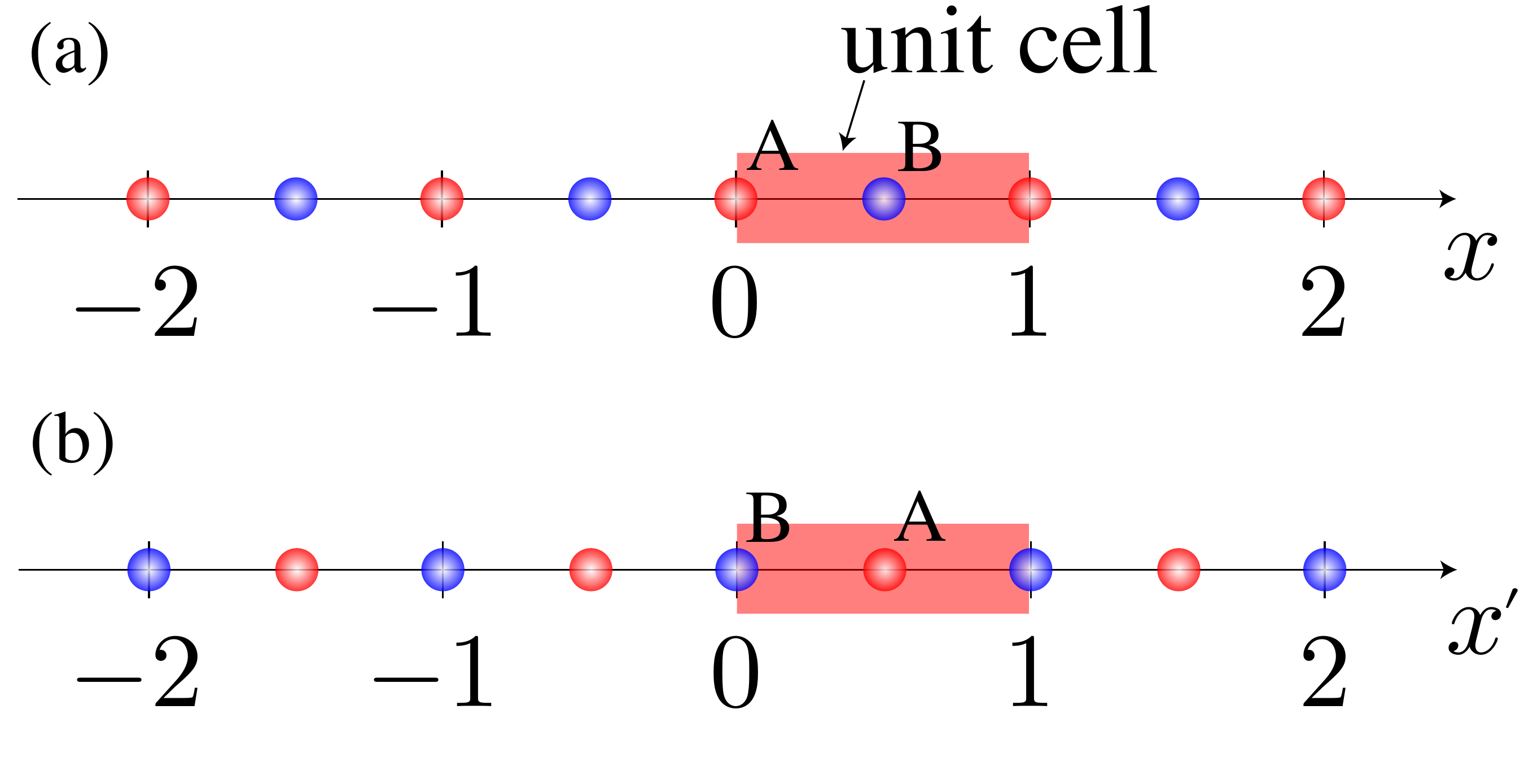}
		\caption{\label{fig:example}Two choices of the coordinate for an inversion symmetric 1D system composed of two atoms A and B in each unit cell. The origin of the coordinate is at the atom A in the panel (a), while it is at the atom B in the panel (b).}
	\end{center}
\end{figure}

To see this subtlety through a simple example, let us consider the inversion symmetric 1D system illustrated in Figure~\ref{fig:example}.  In this model, there are two atoms A and B in a unit cell.
In the panel (a), the atom A is placed at the origin and the coordinate of the atom B is $x=\frac{1}{2}$. 
In the panel (b), the atom B is placed at the origin and the coordinate of the atom A is $x=\frac{1}{2}$. This is the ambiguity of the coordinates mentioned above.
In both cases, the inversion symmetry is about the origin $x=0$, but the atom at the fixed point is different. This means that the inversion in (a) and that in (b) are physically different operations. As a consequence, their representations are not the same and  are related by ${U}_{k_x}^{\alpha}(I)^{\text{(a)}} = e^{i k_x}U_{k_x}^{\alpha}(I)^{\text{(b)}}$.

\subsubsection{The choice of the spin rotation}
As is well-known, the correspondence between a point group element $p_g\in\text{O}(3)$ and the spin rotation matrix $p_{g}^{\text{sp}}\in\text{SU}(2)$ has a sign ambiguity. 
Namely, when $p_{g}^{\text{sp}}\in\text{SU}(2)$ denotes a spin rotation matrix corresponding to the point group $p_g$,  $-p_{g}^{\text{sp}}\in\text{SU}(2)$ is also an equally valid choice of the spin rotation matrix, and there is no unique way of resolving the ambiguity. Our choice of $p_{g}^{\text{sp}}$ is stored in \texttt{srg.dat}.

Given the choice of $p_{g}^{\text{sp}}$ for each $g$, we fix the projective factor in Eq.~\eqref{eq:rep_Gk} by
\begin{align}
	p_{g}^{\text{sp}}p_{g'}^{\text{sp}} &=\omega^{\text{sp}}(g,g')p_{gg'}^{\text{sp}}.
	\label{spinprojective}
\end{align}
The information of $\omega^{\text{sp}}(g,g')$ is stored in \texttt{factor\_system\_spin.dat}.

\section{Installation and usage}
\label{sec4}
\begin{figure}[t]
	\begin{center}
		\includegraphics[width=0.5\columnwidth]{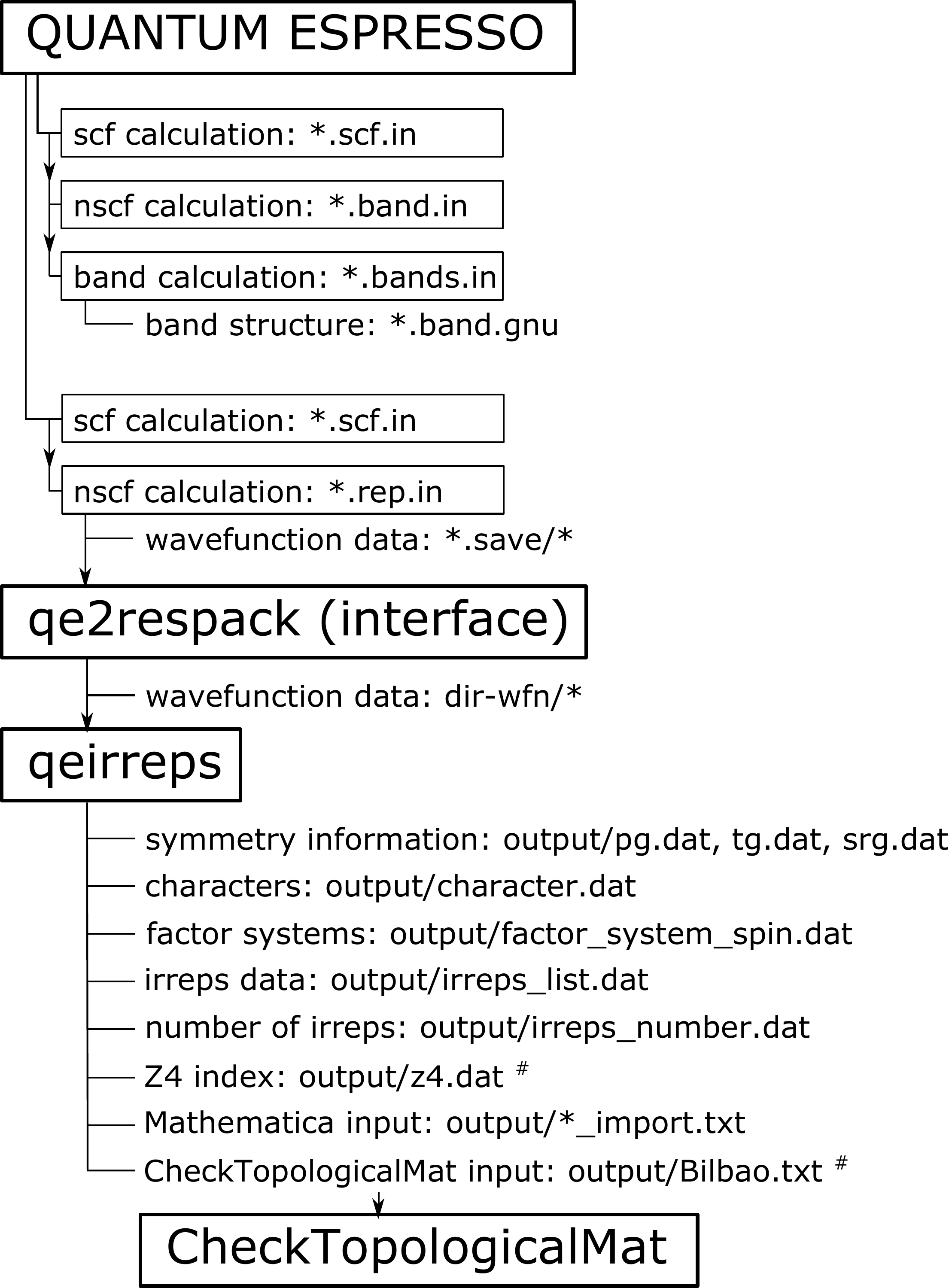}
		\caption{\label{fig:flow}Flowchart of the process to calculate the representation characters of Bloch wavefunctions. 
			The band structure and the wavefunctions of the target material are computed by \textsc{Quantum ESPRESSO}.
			These results are converted to the input files of \texttt{qeirreps} by \texttt{qe2respack}.	
			Finally the representation characters are computed by \texttt{qeirreps}.
			Two files \texttt{z4.dat} and \texttt{Bilbao.txt} (marked by ``\#") are produced only when the options are added to the command. One can use \texttt{Bilbao.txt} as the input file for CheckTopologicalMat~\cite{Vergniory2019}.}
	\end{center}
\end{figure}

Here we explain how to install and use \texttt{qeirreps}. 
The flowchart of the calculations is shown in Figure~\ref{fig:flow}.

\subsection{Compiling environment for \texttt{qeirreps}}
A Fortran 90 compiler, BLAS, and LAPACK libraries are required for the installation of \texttt{qeirreps}. 
\textsc{Quantum ESPRESSO} must also be installed in advance.
In addition, the program \texttt{qe2respack}~\cite{respack}~\footnote{This program \texttt{qe2respack} is a part of the program package RESPACK~\cite{respack}. 
\texttt{qe2respack} in the latest version of RESPACK does not support DFT calculations with spin-orbit coupling. 
For calculations with spin-orbit coupling, we need the specific version of \texttt{qe2respack} described in this section.} is required, which can be downloaded or cloned from the branch of \texttt{respack} ``maxime2" in the GitHub repository (https://github.com/mnmpdadish/respackDev/). The program \texttt{qe2respack.py} is in the directory \texttt{util/qe2respack}.

\subsection{Installation of \texttt{qeirreps}}
Our program \texttt{qeirreps} is released at GitHub (https://github.com/qeirreps/qeirreps). The program files which contain source files, documents, and examples can be cloned or downloaded from this repository. The file \texttt{qeirreps/src/Makefile} must be edited to specify the compiler and libraries in your compiling environment. By typing \texttt{\$ make} in the source directory \texttt{qeirreps/src/}, the executable binary \texttt{qeirreps.x} is compiled. 

\subsection{Preparing input files of \texttt{qeirreps}}\label{sub-sec:prep}
Our program \texttt{qeirreps} works based on the output of \textsc{Quantum ESPRESSO}.
To prepare input files of \texttt{qeirreps}, the following three steps are needed to be done one by one:
\begin{enumerate}
	\item Self-consistent first-principles (scf) calculation of a target material.
	\item Non-self-consistent first-principles (nscf) calculation of the material for each high-symmetry momentum.
	\item Data conversion from \textsc{Quantum ESPRESSO} output files to \texttt{qeirreps} input files.
\end{enumerate}
The first two steps can be carried out by the standard functions of \textsc{Quantum ESPRESSO}.
The result of the scf calculation is used in the successive nscf calculation. 
The set of high-symmetry points to be included in the calculation depends on the purpose of the calculation and the space group of the target material.  When one applies the results of \qeirreps\ to symmetry indicators or topological quantum chemistry, the minimum set of high-symmetry points are listed in Refs.~\cite{QuantitativeMappings, Vergniory2019,Tang2019}.

We refer to the directory that stores the wavefunction data computed by the nscf calculation \texttt{OUTDIR/PREFIX.save} below.
For the moment, \texttt{qeirreps} requires norm-conserving calculations.
Pseudo-potentials must be optimized for norm-conserving calculations (for example, Optimized Norm-Conserving Vanderbilt Pseudopotential~\cite{oncvpsp} is available in PseudoDojo (http://www.pseudo-dojo.org)~\cite{pseudodojo}) and the option \texttt{wf\_collect} should be set \texttt{.TRUE.} in the input files for DFT calculations. 

The output files in \texttt{OUTDIR/PREFIX.save} must be converted by \texttt{qe2respack} into the form of input files of \texttt{qeirreps}. 
The work directory is referred to as \texttt{DIRECTORY\_NAME} here. Then, type the following in the work directory:
\begin{equation*}
\$\ \texttt{python PATH\_OF\_qe2respack/qe2respack.py OUTDIR/PREFIX.save}\ ,
\end{equation*}
where \texttt{PATH\_OF\_qe2respack} is the path to the directory \texttt{util/qe2respack}. \texttt{qeirreps} reads the files produced by \texttt{qe2respack.py} in the \texttt{dir-wfn} directory. The contents of these files are summarized in Table~\ref{tab:qe2respack}.

\begin{table}[htb]
	\caption{\label{tab:qe2respack}List of the output files of \texttt{qe2respack.py}, which serve as the input files of \texttt{qeirreps}.}
	\begin{center}
	\begin{tabular}{|l|l|} \hline
		File name & Information \\ \hline \hline
		\texttt{dat.atom\_position} & The positions of atoms \\ \hline
		\texttt{dat.bandcalc} & The energy cutoff for the wavefunction, Fermi energy, and total energy \\ \hline
		\texttt{dat.sample\_k} & The high-symmetry momenta \\ \hline
		\texttt{dat.eigenvalue} & The energy level of each band \\ \hline
		\texttt{dat.nkm} & The number of reciprocal lattice vectors used in  \\ &the expansion of wavefunctions at each momentum\\ \hline
		\texttt{dat.kg} & The set of reciprocal lattice vectors used in \\ & the expansion of wavefunctions at each momentum\\ \hline
		\texttt{dat.wfn} & The Bloch wavefunction of each band  \\ \hline
		\texttt{dat.lattice} & The lattice vectors \\ \hline
		\texttt{dat.symmetry} &The  symmetry operators \\ \hline		
	\end{tabular}
	\end{center}
\end{table}

\subsection{Running \texttt{qeirreps}}
\label{sub-sec:run}

To run \texttt{qeirreps}, a directory named \texttt{output} must be created in the directory \texttt{DIRECTORY\_NAME} above. Then type
\begin{equation*}
	\texttt{\$ PATH\_OF\_qeirreps/qeirreps.x DIRECTORY\_NAME}\ ,
\end{equation*}
where \texttt{PATH\_OF\_qeirreps} is the path to the directory \texttt{qeirreps/src}.  
There should be 12 output files in \texttt{output}. Seven of them named \texttt{*.dat} contain the following information: (i) the list of the point group part $p_g$ in \texttt{pg.dat}, (ii) the list of the translation part $\bm{t}_g$ in \texttt{tg.dat}, (iii) the list of the spin rotation part $p^{sp}_{g}$ in \texttt{srg.dat}, (iv) the factor system $\omega^{\text{sp}}(g,g')$ associated with the electronic spin in \texttt{factor\_system\_spin.dat}, (v) the representation characters $\chi_{\bm{k}n}(h)$ of Bloch wavefunctions in \texttt{character.dat}, (vi) the characters $\chi_{\bm{k}}^\alpha$ of irreducible representations of $\calGk$ in \texttt{irreps\_list.dat}, and (vii) the numbers $n_{\bm{k}n}^\alpha$ of irreducible representations of $\calGk$ in the $n$-th energy level in \texttt{irreps\_number.dat}. The contents of these files are summarized in Table~\ref{tab:qeirreps}.
Five of them named \texttt{*\_import.txt} are the corresponding files for Mathematica usage.
The standard output that appears during this calculation shows the lattice vectors, reciprocal lattice vectors, operation type of each symmetry operators, and so on. 

For materials with inversion symmetry, \texttt{qeirreps} also implements the automatic evaluation of the $\mZ_4$ index. To do this, an option of filling should be added to the command as 
\begin{equation*}
	\texttt{\$ PATH\_OF\_qeirreps/qeirreps.x DIRECTORY\_NAME FILLING z4} .
\end{equation*}
Here, \texttt{z4} is the option to obtain the $\mZ_4$ index and \texttt{FILLING} is the number of electrons per unit cell of the target material and is shown in the standard output of scf calculation by \textsc{Quantum ESPRESSO} as ``\texttt{number of electrons = FILLING}." Then, the program generates an additional output named \texttt{z4.dat} that contains the value of the $\mZ_4$ index.

Our program \texttt{qeirreps} also provides the input file for the tool CheckTopologicalMat (https://www.cryst.ehu.es/cgi-bin/cryst/programs/topological.pl)~\cite{Vergniory2019}. 
Our program exports the input file named \texttt{Bilbao.txt} with the following command
\begin{equation*}
	\texttt{\$ PATH\_OF\_qeirreps/qeirreps.x DIRECTORY\_NAME FILLING ctm} ,
\end{equation*} 
where \texttt{ctm} is the option to obtain the input file. The tool CheckTopologicalMat tells us topological properties such as the value of symmetry indicators and gaplessness.

\begin{table}[htb]
	\caption{\label{tab:qeirreps} List of the output files of \texttt{qeirreps}. Two files \texttt{z4.dat} and \texttt{Bilbao.txt} (marked by ``\#") are produced only when the options are added to the command. The files named \texttt{*\_import.txt} are for the Mathematica usage.}
	\begin{center}
	\begin{tabular}{|l|l|} \hline
		File name & Information \\ \hline \hline
		\texttt{pg.dat} & The point group part ($p_g$) of each symmetry operation \\ \hline
		\texttt{tg.dat} &  The translation part ($t_g$) of each symmetry operation \\ \hline
		\texttt{srg.dat} & The spin rotation part ($p_g^{\mathrm{sp}}$) of each symmetry operation \\ \hline
		\texttt{factor\_system\_spin.dat} & The factor system ($\omega^{\mathrm{sp}}(g,g')$) associated with the electric spin \\ \hline
		\texttt{character.dat} & The character tables of irreducible representations of Bloch wavefunctions\\ \hline
		\texttt{irreps\_list.dat} & The characters ($\chi_{\bm{k}}^\alpha$) of irreducible representations of $\calGk$\\ \hline
		\texttt{irreps\_number.dat} & The numbers ($n_{\bm{k}n}^\alpha$) of irreducible representations of $\calGk$ in the $n$-th energy level \\ \hline
		\texttt{z4.dat}$^\#$ & The $\mZ_4$ index \\ \hline
		\texttt{Bilbao.txt}$^\#$ & The input file for CheckTopologicalMat~\cite{Vergniory2019} \\ \hline		
	\end{tabular}
	\end{center}
\end{table}

\section{Examples}\label{sec:ex}

In this section, we discuss several examples to demonstrate the usage of \texttt{qeirreps}.
We also explain how to compute topological indices based on the output of \texttt{qeirreps}.
In all the examples, we use the primitive unit cell in the calculation of the band structure and the space group representations in the band structure.

\subsection{Bismuth}\label{sub-sec:ex-bi}
\begin{figure}[H]
	\begin{center}
		\includegraphics[width=0.75\columnwidth]{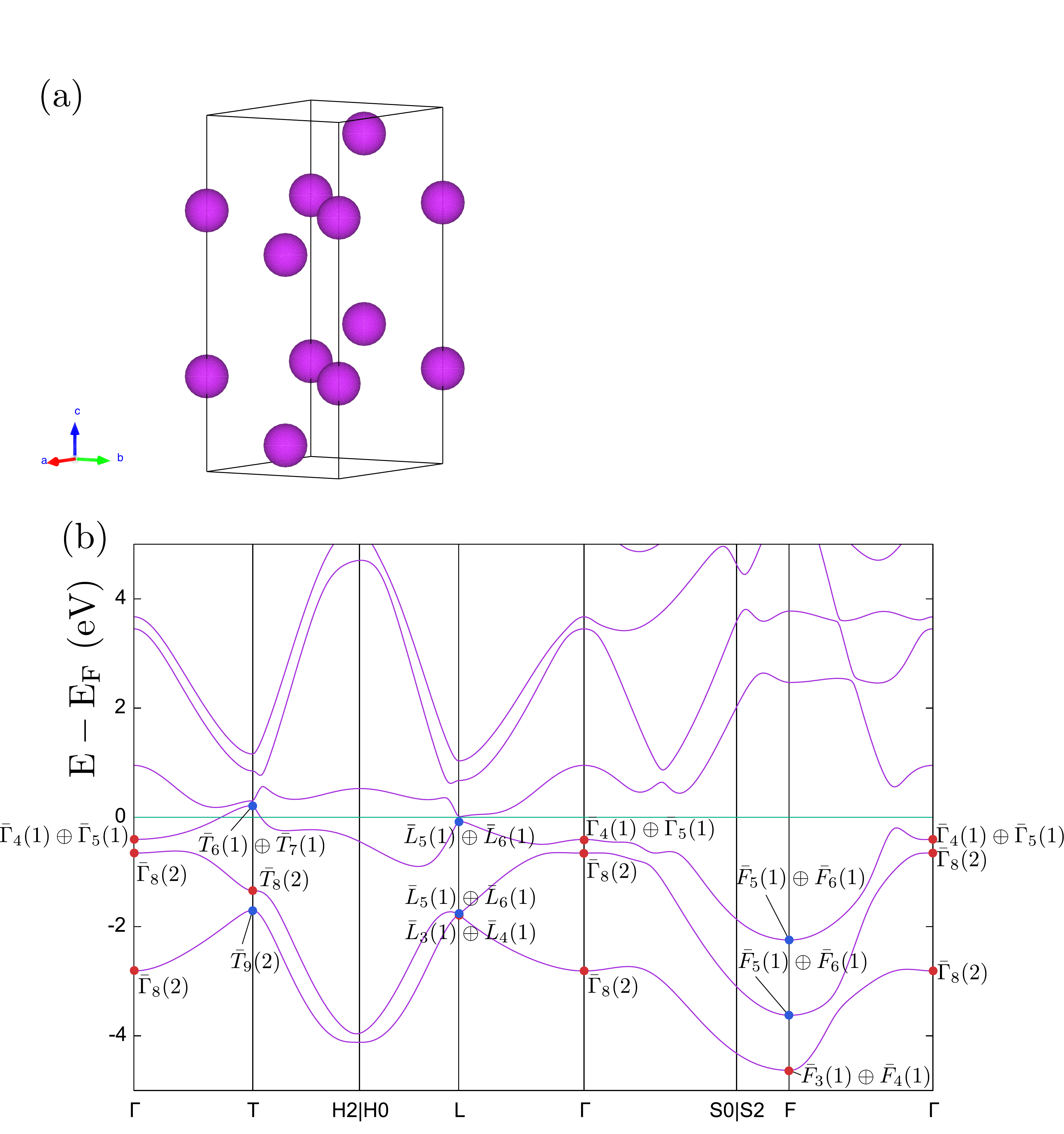}
		\caption{\label{fig:bismuth}
(a) The crystal structure~\cite{mp} and the Brillouin zone~\cite{seekpath1,seekpath2} of bismuth. (b) The band structure of bismuth with spin-orbit coupling.\protect \footnotemark $\bar{K}_{\alpha}(m)$ denotes a $m$-dimensional irreducible representation $U_{K}^{\alpha}$ at a high-symmetry point $K$. The correspondence between these labels and representation characters is included in~\ref{app:ch-bi}. The color of dots represents the inversion parity: when $\bar{K}_{\alpha}(m)$ is marked red (blue), all $m$ levels have the even (odd) parity.
}
	\end{center}
\end{figure}

Our first example is bismuth. The space group ${R\bar{3}m}$ (No.$166$) contains the inversion symmetry $I$. 
We obtain the crystal structure data of bismuth [Figure~\ref{fig:bismuth} (a)] from Material Project~\cite{mp} and converted it into the form of an input file for \textsc{Quantum ESPRESSO} (included in~\ref{app:bi-scf-in}) by SeeK-path~\cite{seekpath1,seekpath2}~\footnote{In this particular example, we manually shifted positions of atoms in the primitive unit cell so that we can compare our results with the one in Ref.~\cite{Vergniory2019}}. 

We compute the irreducible representations of wavefunctions by \texttt{qeirreps}, taking into account the spin-orbit coupling.
Our results are shown in the band structure in Figure~\ref{fig:bismuth} (b).
The correspondence between the labels and characters of irreducible representations is included in~\ref{app:ch-bi}. 
These results are consistent with previous studies~\cite{Vergniory2019,Bilbao}. 

We also compute the $\mZ_4$ index~\cite{PhysRevX.8.031070, QuantitativeMappings} using the option explained in Sec.~\ref{sub-sec:run}. The output (\texttt{z4.dat}) shows
\begin{eqnarray*}
	&&\texttt{sum of parities for           8 k-points:}\\
	&&\texttt{  -7.99999999999963}\\
	&&\texttt{ z4 index:}\\
	&&\texttt{ 2}
\end{eqnarray*}
Namely, the $\mZ_4$ index for bismuth with  significant spin-orbit coupling is $2$, indicating that this material is a higher-order topological insulator~\cite{Schindler:2018aa,Tang2019,Vergniory2019,Hsu13255}. The sample input files for bismuth are available in the directory \texttt{qeirreps/example}. 
\subsection{Silicon}
\label{sub-sec:ex-si}
\begin{figure}[H]
	\begin{center}
		\includegraphics[width=0.75\columnwidth]{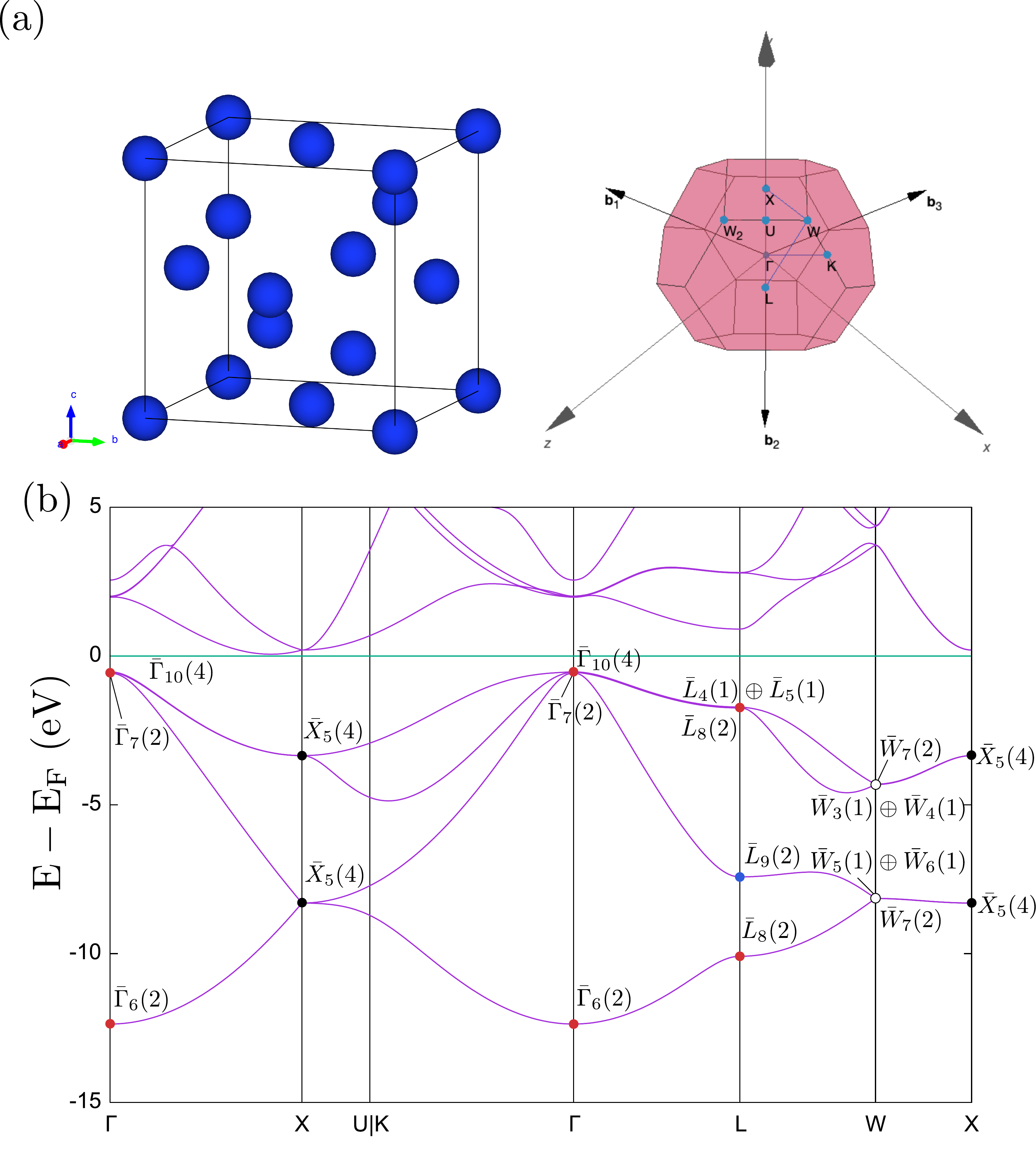}
		\caption{\label{fig:silicon}
(a) The crystal structure~\cite{mp} and the Brillouin zone~\cite{seekpath1,seekpath2} of silicon. (b) The band structure of silicon with spin-orbit coupling. $\bar{K}_{\alpha}(m)$ denotes a $m$-dimensional irreducible representation $U_{K}^{\alpha}$ at a high-symmetry point $K$. The correspondence between these labels and representation characters is included in~\ref{app:ch-si}. The color of dots represents the inversion parity: when $\bar{K}_{\alpha}(m)$ is marked red (blue), all $m$ levels have the even (odd) parity, and when it is black, the half of them are even and the other half is odd. The little group at $W$ does not have the inversion symmetry and open circles represent the absence of the inversion symmetry.}
	\end{center}
\end{figure}

To demonstrate that \texttt{qeirreps} equally works for nonsymmorphic space groups, let us discuss silicon. Its space group is ${Fd\bar{3}m}$ (No. $227$), which also contains the inversion symmetry $I$. 
The calculation procedure is completely the same as those for bismuth in Sec.~\ref{sub-sec:ex-bi}. 
Our results of irreducible representations are in Figure~\ref{fig:silicon} (b). The $\mZ_4$ index is found to be zero. These results are consistent with the previous study in Ref.~\cite{Vergniory2019}. Sample input files for silicon are also in the directory \texttt{qeirreps/example}.


\footnotetext{``H0” and ``H2” are distinct momenta (i.e., their difference is not a reciprocal lattice vector) but they are related by a symmetry operation. The same is true for “S0” and “S2”.}

\subsection{NaCdAs}
\label{sub-sec:ex-NaCdAs}
\begin{figure}[H]
	\begin{center}
		\includegraphics[width=0.75\columnwidth]{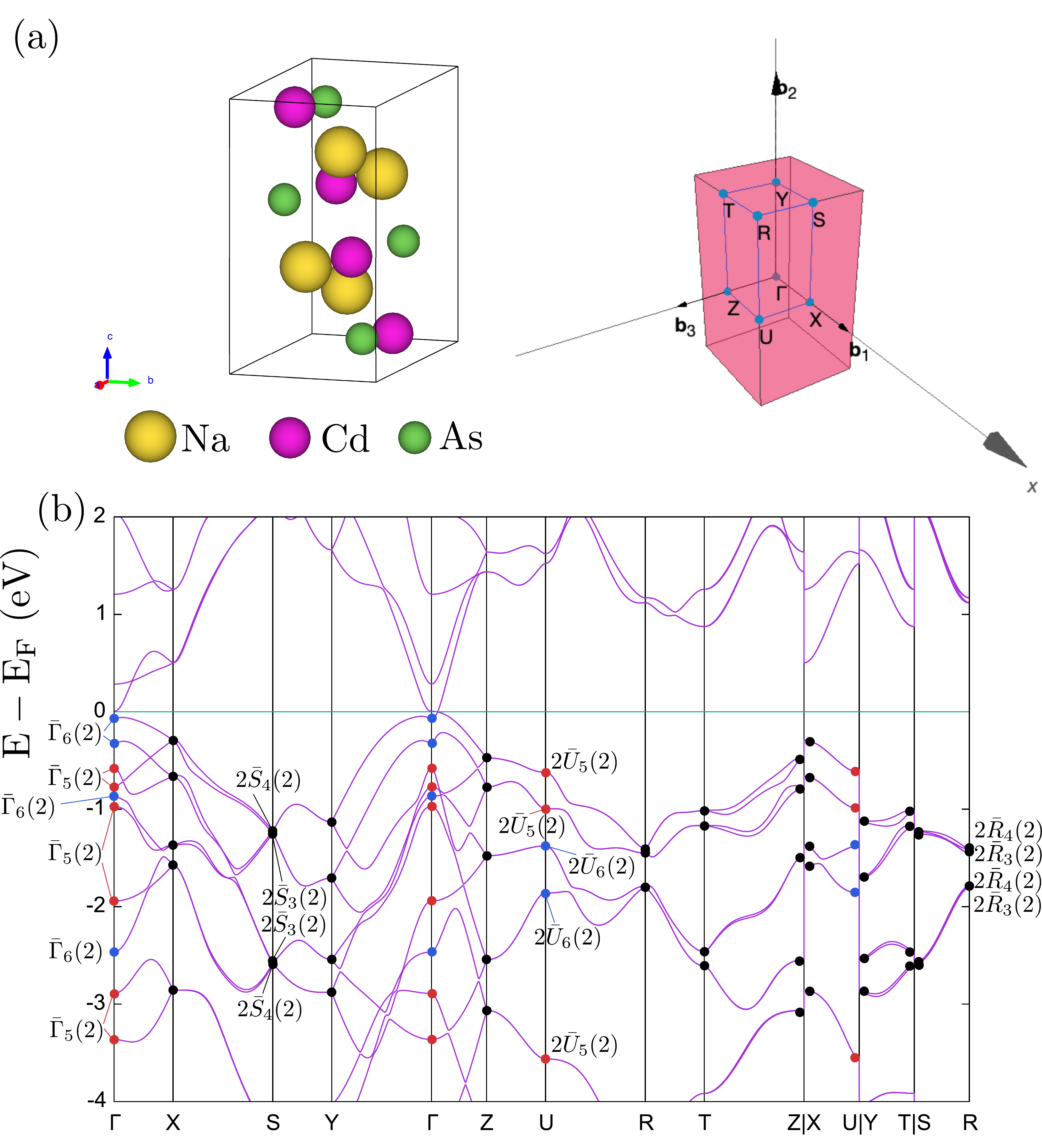}
		\caption{\label{fig:NaCdAs}
(a) The crystal structure~\cite{mp} and the Brillouin zone~\cite{seekpath1,seekpath2} of NaCdAs. (b) The band structure of NaCdAs with spin-orbit coupling. $\bar{K}_{\alpha}(m)$ denotes a $m$-dimensional irreducible representation $U_{K}^{\alpha}$ at a high-symmetry point $K$. The correspondence between these labels and representation characters is included in~\ref{app:ch-NaCdAs}. The color of dots represents the inversion parity: when $\bar{K}_{\alpha}(m)$ is marked red (blue), all $m$ levels have the even (odd) parity, and when it is black, the half of them are even and the other half is odd.}
	\end{center}
\end{figure}

Here we discuss NaCdAs as an example of topological materials with nonsymmorphic space group symmetries. Its space group is $Pnma$ (No. $62$) which has the inversion symmetry $I$. Irreducible representations computed in the presence of spin-orbit coupling are shown in Figure~\ref{fig:NaCdAs} (b). The $\mZ_4$ index is 1, indicating that this material is a candidate of the strong $\mathbb{Z}_2$ topological insulator. These results are consistent with the previous study in Ref.~\cite{Vergniory2019}. (The $\mZ_4$ index in Ref.~\cite{Vergniory2019} is 3 because their definition of the index includes an additional minus sign.) Sample input files for NaCdAs are contained in the directory \texttt{qeirreps/example}.

\subsection{PbPt$_3$}
\label{sub-sec:ex-PbPt3}
\begin{figure}[H]
	\begin{center}
		\includegraphics[width=0.75\columnwidth]{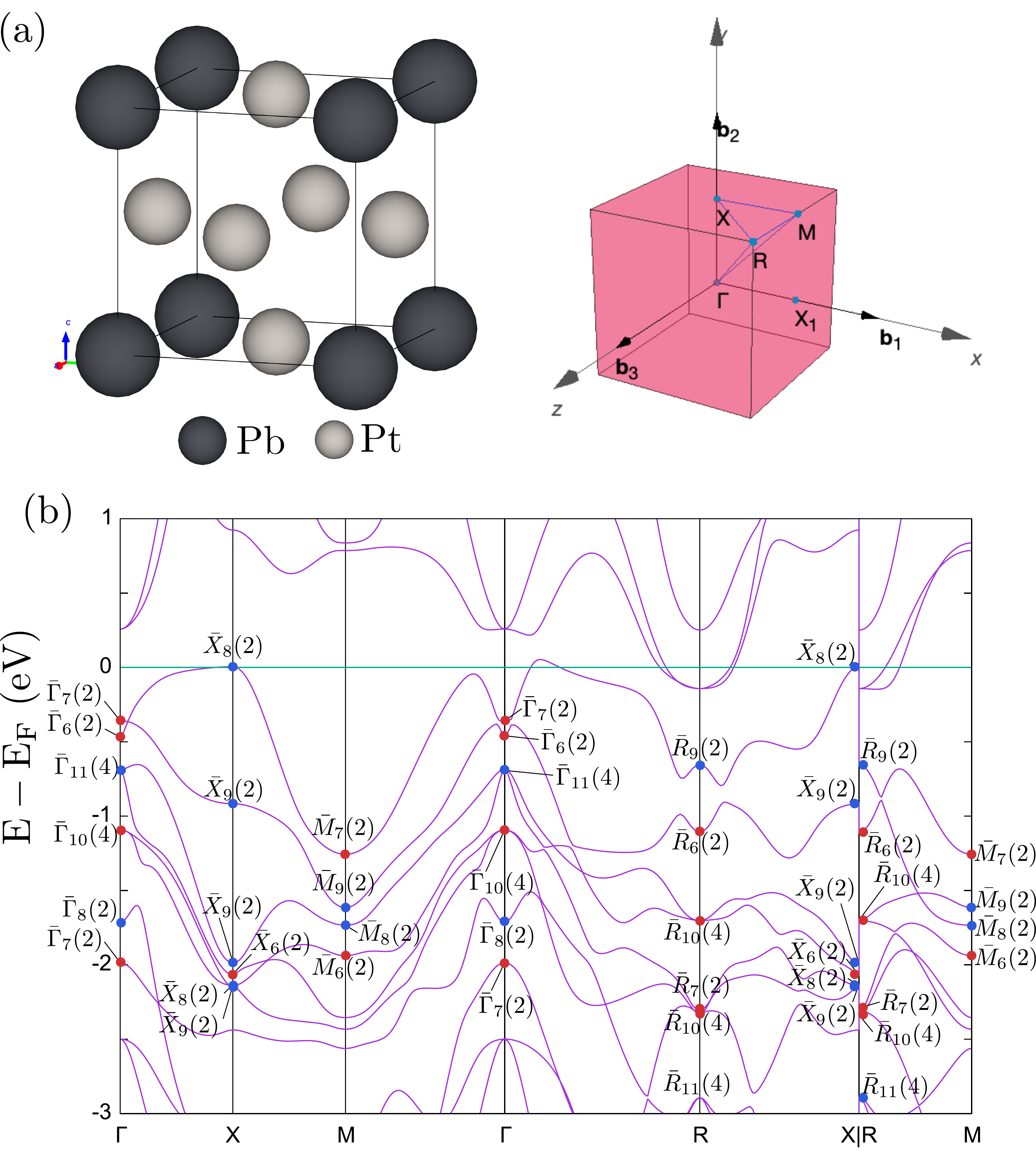}
		\caption{\label{fig:PbPt3}
(a) The crystal structure~\cite{mp} and the Brillouin zone~\cite{seekpath1,seekpath2} of $\mathrm{PbPt_3}$. (b) The band structure of $\mathrm{PbPt_3}$ with spin-orbit coupling. $\bar{K}_{\alpha}(m)$ denotes a $m$-dimensional irreducible representation $U_{K}^{\alpha}$ at a high-symmetry point $K$. The correspondence between these labels and representation characters is included in~\ref{app:ch-pbpt3}. The color of dots represents the inversion parity: when $\bar{K}_{\alpha}(m)$ is marked red (blue), all $m$ levels have the even (odd) parity.}
	\end{center}
\end{figure}

As our fourth example, let us discuss PbPt$_3$, whose space group ${Pm\bar{3}m}$ (No. $221$).  This space group contains various elements such as the inversion $(I)$, the rotoinversion about the $z$-axis $(S_{4}^z)$, and the mirror symmetries. We also assume the time-reversal symmetry. Our results of irreducible representations in the presence of spin-orbit coupling are in Figure~\ref{fig:PbPt3} (b).

In this symmetry settings, we can define a $\mZ_4$ index (other than the sum of the inversion parities) and a $\mZ_8$ index by~\cite{PhysRevB.86.115112, PhysRevX.8.031070,QuantitativeMappings}
\begin{align}
z_4 &=\frac{3}{2}n_{R}^{7}-\frac{3}{2}n_{R}^{9}-\frac{1}{2}n_{R}^{6}+\frac{1}{2}n_{R}^{8}+n_{R}^{10}-n_{R}^{11} + \frac{3}{2}n_{X}^{7}-\frac{3}{2}n_{X}^{9}-\frac{1}{2}n_{X}^{6}+\frac{1}{2}n_{X}^{8} \nonumber\\
&\quad \quad \quad\quad \quad \quad\quad \quad \quad\quad \quad \quad\quad \quad \quad\quad \quad \quad\quad \quad \quad+n_{M}^{6}+n_{M}^{7}-n_{M}^{8}-n_{M}^{9}\mod 4,\\
z_8&=\frac{1}{4}\sum_{K\in \text{TRIMs}}\sum_{\alpha_{K}}\chi_{K}^{\alpha_K}(I) n_{K}^{\alpha_K} - \frac{1}{\sqrt{2}}\sum_{K'\in K_4}\sum_{\alpha_{K'}}\chi_{K'}^{\alpha_{K'}}(S_{4}^z) n_{K'}^{\alpha_{K'}} \mod 8,
\end{align}
where $n_{K}^{\alpha_K}$ represents the number of occupied states that have irreducible representations $U_{K}^{\alpha_K}$ of $\calG_K$, and the set of $\{n_{K}^{\alpha_K}\}$ is included in \ref{app:b-pbpt3}. $K_4$ denotes four momenta invariant under the $S_{4}^z$.
Although our program \texttt{qeirreps}  does not offer an automated calculation of these indices, one can compute them manually by using \texttt{character.dat} or uploading \texttt{Bilbao.txt} to CheckTopologicalMat~\cite{Vergniory2019}. In this example, we find $(z_4, z_8) = (3,6)$, which is consistent with Ref.~\cite{Vergniory2019}. According to Ref.~\cite{PhysRevX.8.031070,QuantitativeMappings}, these values indicate nontrivial mirror Chern numbers. Sample input files for PbPt$_3$ are included in \texttt{qeirreps/example}.

\section{Conclusions}\label{sec5}
In conclusion, we developed a new open-source code \texttt{qeirreps} for computing irreducible representations in band structures based on the output of \textsc{Quantum ESPRESSO}. 
We explained the installation of the program and demonstrated its usage through examples. 
When combined with the symmetry indicator method, the output of this program can be used to diagnose the topological property of weakly interacting materials.
Since \textsc{Quantum ESPRESSO} is a free, widely-used software,
\texttt{qeirreps} should accelerate the exploration of new topological materials.

\section*{Acknowledgement}
We thank Ryotaro Arita and Motoaki Hirayama for frutuful discussions. We also thank Luis Elcoro for useful and kind correspondence. We are very grateful to the anonymous referee of this manuscript for his or her constructive comments and suggestions. All DFT calculations in this work has been done using the facilities of the Supercomputer Center, the Institute for Solid State Physics, the University of Tokyo.
The work of AM and SO is supported by Materials Education program for the future leaders in Research, Industry, and Technology (MERIT). 
SO is also supported by the ANRI Fellowship and JSPS KAKENHI Grant No.~JP20J21692.
The work of YN is supported by JSPS KAKENHI Grant No.~16H06345, 17K14336, 18H01158, 20K14423. 
The work of HW is supported by JSPS KAKENHI Grant No.~JP17K17678 and by JST PRESTO Grant No.~JPMJPR18LA.

\appendix

\section{Samples of \textsc{Quantum ESPRESSO} input}

\subsection{For the scf calculation of bismuth}\label{app:bi-scf-in}
\begin{lstlisting}[basicstyle=\ttfamily\footnotesize]
&CONTROL
    calculation = 'scf'
    restart_mode='from_scratch',
    prefix='Bi'
    pseudo_dir='../../pseudo/soc/'
    outdir='./work/scf/'
    tstress=.true.
    tprnfor=.true.
    wf_collect=.true.
/
&SYSTEM
    ibrav = 0
    nat = 2
    ntyp = 1
    ecutwfc=100.
    occupations='smearing'
    smearing='m-p'
    degauss=0.01
    use_all_frac=.true.
    lspinorb=.true.
    noncolin=.true.
/
&ELECTRONS
    mixing_beta=0.3
    conv_thr=1.0d-8
/
ATOMIC_SPECIES
Bi	208.9804	Bi.upf
ATOMIC_POSITIONS angstrom
Bi  2.30479336  1.330673067  1.153637296
Bi -0.00000000 -0.000000000  2.838187257
K_POINTS automatic
8 8 8 0 0 0
CELL_PARAMETERS angstrom
    2.3047933600     1.3306730668     3.9918245533
   -2.3047933600     1.3306730668     3.9918245533
    0.0000000000    -2.6613461336     3.9918245533
\end{lstlisting}

\subsection{For the nscf calculation of bismuth}\label{app:bi-rep-in}
\begin{lstlisting}[basicstyle=\ttfamily\footnotesize]
&CONTROL
    calculation = 'bands'
    restart_mode='from_scratch',
    prefix='Bi'
    pseudo_dir='../../pseudo/soc/'
    outdir='./work/rep/'
    tstress=.true.
    tprnfor=.true.
    wf_collect=.true.
/
&SYSTEM
    ibrav = 0
    nat = 2
    ntyp = 1
    ecutwfc=100.
    occupations='smearing'
    smearing='m-p'
    degauss=0.01
    use_all_frac=.true.
    lspinorb=.true.
    noncolin=.true.
/
&ELECTRONS
    mixing_beta=0.3
    conv_thr=1.0d-8
/
ATOMIC_SPECIES
Bi	208.9804	Bi.upf
ATOMIC_POSITIONS angstrom
Bi  2.30479336  1.330673067  1.153637296
Bi -0.00000000 -0.000000000  2.838187257
K_POINTS crystal
8
0.0	0.0	0.0	1
0.5	0.0	0.0	1
0.0	0.5	0.0	1
0.0	0.0	0.5	1
0.0	0.5	0.5	1
0.5	0.0	0.5	1
0.5	0.5	0.0	1
0.5	0.5	0.5	1
CELL_PARAMETERS angstrom
    2.3047933600     1.3306730668     3.9918245533
   -2.3047933600     1.3306730668     3.9918245533
    0.0000000000    -2.6613461336     3.9918245533
\end{lstlisting}

\subsection{For the scf calculation of silicon}\label{app:si-scf-in}
\begin{lstlisting}[basicstyle=\ttfamily\footnotesize]
&CONTROL
    calculation = 'scf'
    restart_mode='from_scratch',
    prefix='Si'
    pseudo_dir='../../pseudo/soc'
    outdir='./work/scf/'
    tstress=.true.
    tprnfor=.true.
    wf_collect=.true.
/
&SYSTEM
    ibrav = 0
    nat = 2
    ntyp = 1
    ecutwfc=100.
    occupations='smearing'
    smearing='m-p'
    degauss=0.01
    use_all_frac=.true.
    lspinorb=.true.
    noncolin=.true.
/
&ELECTRONS
    mixing_beta=0.3
    conv_thr=1.0d-8
/
ATOMIC_SPECIES
Si	28.0855	Si.upf
ATOMIC_POSITIONS crystal
Si       0.625     0.625     0.625
Si       0.375     0.375     0.375
K_POINTS automatic
8 8 8 0 0 0
CELL_PARAMETERS angstrom
    0.0000000000     2.7347724300     2.7347724300
    2.7347724300     0.0000000000     2.7347724300
    2.7347724300     2.7347724300     0.0000000000
\end{lstlisting}

\subsection{For the nscf calculation of silicon}\label{app:si-rep-in}
\begin{lstlisting}[basicstyle=\ttfamily\footnotesize]
&CONTROL
    calculation = 'bands'
    restart_mode='from_scratch',
    prefix='Si'
    pseudo_dir='../../pseudo/soc'
    outdir='./work/rep/'
    tstress=.true.
    tprnfor=.true.
    wf_collect=.true.
/
&SYSTEM
    ibrav = 0
    nat = 2
    ntyp = 1
    ecutwfc=100.
    occupations='smearing'
    smearing='m-p'
    degauss=0.01
    use_all_frac=.true.
    lspinorb=.true.
    noncolin=.true.
/
&ELECTRONS
    mixing_beta=0.3
    conv_thr=1.0d-8
/
ATOMIC_SPECIES
Si	28.0855	Si.upf
ATOMIC_POSITIONS crystal
Si       0.625     0.625     0.625
Si       0.375     0.375     0.375
K_POINTS crystal
8
0.0	0.0	0.0	1
0.5	0.0	0.0	1
0.0	0.5	0.0	1
0.0	0.0	0.5	1
0.0	0.5	0.5	1
0.5	0.0	0.5	1
0.5	0.5	0.0	1
0.5	0.5	0.5	1
CELL_PARAMETERS angstrom
    0.0000000000     2.7347724300     2.7347724300
    2.7347724300     0.0000000000     2.7347724300
    2.7347724300     2.7347724300     0.0000000000
\end{lstlisting}

\subsection{For the scf calculation of NaCdAs}\label{app:NaCdAs-scf-in}
\begin{lstlisting}[basicstyle=\ttfamily\footnotesize]
&CONTROL
    calculation = 'scf'
    restart_mode='from_scratch',
    prefix='NaCdAs'
    pseudo_dir='../../pseudo/soc'
    outdir='./work/scf'
    tstress=.true.
    tprnfor=.true.
    wf_collect=.true.
/
&SYSTEM
    ibrav = 0
    nat = 12
    ntyp = 3
    ecutwfc=100.
    occupations='smearing'
    smearing='m-p'
    degauss=0.01
    use_all_frac=.true.
    lspinorb=.true.
    noncolin=.true.
/
&ELECTRONS
    mixing_beta=0.3
    conv_thr=1.0d-8
/
ATOMIC_SPECIES
Na      22.989769       Na.upf
Cd      112.411         Cd.upf
As      74.9216         As.upf
ATOMIC_POSITIONS crystal
Na      0.0184820000    0.2500000000    0.1734590000
Na      0.9815180000    0.7500000000    0.8265410000
Na      0.4815180000    0.7500000000    0.6734590000
Na      0.5184820000    0.2500000000    0.3265410000
Cd      0.1500210000    0.2500000000    0.5744130000
Cd      0.8499790000    0.7500000000    0.4255870000
Cd      0.3499790000    0.7500000000    0.0744130000
Cd      0.6500210000    0.2500000000    0.9255870000
As      0.2214520000    0.7500000000    0.3907850000
As      0.7785480000    0.2500000000    0.6092150000
As      0.2785480000    0.2500000000    0.8907850000
As      0.7214520000    0.7500000000    0.1092150000
K_POINTS automatic
8 8 8 0 0 0
CELL_PARAMETERS angstrom
    7.6494810000     0.0000000000     0.0000000000
    0.0000000000     4.5188670000     0.0000000000
    0.0000000000     0.0000000000     8.1790120000
    
\end{lstlisting}

\subsection{For the nscf calculation of NaCdAs}\label{app:NaCdAs-rep-in}
\begin{lstlisting}[basicstyle=\ttfamily\footnotesize]
&CONTROL
    calculation = 'bands'
    restart_mode='from_scratch',
    prefix='NaCdAs'
    pseudo_dir='../../pseudo/soc'
    outdir='./work/rep'
    tstress=.true.
    tprnfor=.true.
    wf_collect=.true.
/
&SYSTEM
    ibrav = 0
    nat = 12
    ntyp = 3
    ecutwfc=100.
    occupations='smearing'
    smearing='m-p'
    degauss=0.01
    use_all_frac=.true.
    lspinorb=.true.
    noncolin=.true.
/
&ELECTRONS
    mixing_beta=0.3
    conv_thr=1.0d-8
/
ATOMIC_SPECIES
Na      22.989769       Na.upf
Cd      112.411         Cd.upf
As      74.9216         As.upf
ATOMIC_POSITIONS crystal
Na      0.0184820000    0.2500000000    0.1734590000
Na      0.9815180000    0.7500000000    0.8265410000
Na      0.4815180000    0.7500000000    0.6734590000
Na      0.5184820000    0.2500000000    0.3265410000
Cd      0.1500210000    0.2500000000    0.5744130000
Cd      0.8499790000    0.7500000000    0.4255870000
Cd      0.3499790000    0.7500000000    0.0744130000
Cd      0.6500210000    0.2500000000    0.9255870000
As      0.2214520000    0.7500000000    0.3907850000
As      0.7785480000    0.2500000000    0.6092150000
As      0.2785480000    0.2500000000    0.8907850000
As      0.7214520000    0.7500000000    0.1092150000
K_POINTS crystal
8
0.0     0.0     0.0     1
0.5     0.0     0.0     1
0.0     0.5     0.0     1
0.0     0.0     0.5     1
0.0     0.5     0.5     1
0.5     0.0     0.5     1
0.5     0.5     0.0     1
0.5     0.5     0.5     1
CELL_PARAMETERS angstrom
    7.6494810000     0.0000000000     0.0000000000
    0.0000000000     4.5188670000     0.0000000000
    0.0000000000     0.0000000000     8.1790120000
\end{lstlisting}

\subsection{For the scf calculation of $PbPt_3$}\label{app:pbpt3-scf-in}
\begin{lstlisting}[basicstyle=\ttfamily\footnotesize]
&CONTROL
    calculation = 'scf'
    restart_mode='from_scratch',
    prefix='PbPt3'
    pseudo_dir='../../pseudo/soc/'
    outdir='./work/scf'
    tstress=.true.
    tprnfor=.true.
    wf_collect=.true.
/
&SYSTEM
    ibrav = 0
    nat = 4
    ntyp = 2
    ecutwfc=100.
    occupations='smearing'
    smearing='m-p'
    degauss=0.01
    use_all_frac=.true.
    lspinorb=.true.
    noncolin=.true.
/
&ELECTRONS
    mixing_beta=0.3
    conv_thr=1.0d-8
/
ATOMIC_SPECIES
Pt      195.084 Pt.upf
Pb      207.2   Pb.upf
ATOMIC_POSITIONS angstrom
Pt       0.0000000000     2.0660527150     2.0660527150
Pt       2.0660527150     0.0000000000     2.0660527150
Pt       2.0660527150     2.0660527150     0.0000000000
Pb       0.0000000000     0.0000000000     0.0000000000
K_POINTS automatic
8 8 8 0 0 0
CELL_PARAMETERS angstrom
    4.1321054300     0.0000000000     0.0000000000
    0.0000000000     4.1321054300     0.0000000000
    0.0000000000     0.0000000000     4.1321054300
\end{lstlisting}

\subsection{For the nscf calculation of $PbPt_3$}\label{app:pbpt3-rep-in}
\begin{lstlisting}[basicstyle=\ttfamily\footnotesize]
&CONTROL
    calculation = 'bands'
    restart_mode='from_scratch',
    prefix='PbPt3'
    pseudo_dir='../../pseudo/soc/'
    outdir='./work/rep'
    tstress=.true.
    tprnfor=.true.
    wf_collect=.true.
/
&SYSTEM
    ibrav = 0
    nat = 4
    ntyp = 2
    ecutwfc=100.
    occupations='smearing'
    smearing='m-p'
    degauss=0.01
    use_all_frac=.true.
    lspinorb=.true.
    noncolin=.true.
/
&ELECTRONS
    mixing_beta=0.3
    conv_thr=1.0d-8
/
ATOMIC_SPECIES
Pt      195.084 Pt.upf
Pb      207.2   Pb.upf
ATOMIC_POSITIONS angstrom
Pt       0.0000000000     2.0660527150     2.0660527150
Pt       2.0660527150     0.0000000000     2.0660527150
Pt       2.0660527150     2.0660527150     0.0000000000
Pb       0.0000000000     0.0000000000     0.0000000000
K_POINTS crystal
4
0.0     0.0     0.0     1
0.0     0.5     0.0     1
0.5     0.5     0.0     1
0.5     0.5     0.5     1
CELL_PARAMETERS angstrom
    4.1321054300     0.0000000000     0.0000000000
    0.0000000000     4.1321054300     0.0000000000
    0.0000000000     0.0000000000     4.1321054300
    
\end{lstlisting}

\section{Information of irreducible representations}\label{app:ch}
The notations of irreducible representations in the following tables follow Ref.~\cite{Bilbao}. Note that some characters have different values due to the choice of fractional translations $\bt_g$ and spin rotation matrices $p_g^{\text{sp}}$.
\subsection{Character tables for bismuth~\cite{Bilbao}}\label{app:ch-bi}
\restylefloat{table}
\begin{table}[H]
\begin{tabular}{|c|c||c|c|c|c|c|c|} 
		\hline
		$p_g$ & $p_g^{\text{sp}}$ & $\bar{\Gamma}_4$ & $\bar{\Gamma}_5$ & $\bar{\Gamma}_6$ & $\bar{\Gamma}_7$ & $\bar{\Gamma}_8$ & $\bar{\Gamma}_9$  \\ \hline \hline
		$1$ & 							$\left(\begin{array}{cc} 1&0 \\ 0&1 \end{array}\right)$ & $1$ &  $1$ &  $1$ &  $1$ &  $2$ &  $2$ \\ \hline
		$3^{+}_{001}$ &				$\left(\begin{array}{cc} (1+\sqrt{3}i)/2&0 \\ 0&(1-\sqrt{3}i)/2 \end{array}\right)$ & $-1$ & $-1$ & $-1$ & $-1$ & $1$ & $1$ \\ \hline
		$3^{-}_{001}$ &				$\left(\begin{array}{cc} (1-\sqrt{3}i)/2&0 \\ 0&(1+\sqrt{3}i)/2 \end{array}\right)$ &$-1$ & $-1$ & $-1$ & $-1$ & $1$ & $1$ \\ \hline
		$2_{1\bar{1}0}$ &					$\left(\begin{array}{cc} 0&i \\ i&0 \end{array}\right)$ & $-i$ & $i$ & $-i$ & $i$ & $0$ & $0$ \\ \hline
		$2_{01\bar{1}}$ &				$\left(\begin{array}{cc} 0&-(\sqrt{3}-i)/2 \\ (\sqrt{3}+i)/2&0 \end{array}\right)$ & $i$ & $-i$ & $i$ & $-i$ & $0$ & $0$ \\ \hline
		$2_{\bar{1}01}$ &		$\left(\begin{array}{cc} 0&-(\sqrt{3}+i)/2 \\ (\sqrt{3}-i)/2&0 \end{array}\right)$ & $-i$ & $i$ & $-i$ & $i$ & $0$ & $0$ \\ \hline
		$I$ & 					$\left(\begin{array}{cc} 1&0 \\ 0&1 \end{array}\right)$ & $1$ & $1$ & $-1$ & $-1$ & $2$ & $-2$ \\ \hline
		$I3^{+}_{001}$ &		$\left(\begin{array}{cc} (1+\sqrt{3}i)/2&0 \\ 0&(1-\sqrt{3}i)/2 \end{array}\right)$ &$-1$ & $-1$ & $1$ & $1$ & $1$ & $-1$ \\ \hline
		$I3^{-}_{001}$ &		$\left(\begin{array}{cc} (1-\sqrt{3}i)/2&0 \\ 0&(1+\sqrt{3}i)/2 \end{array}\right)$ &$-1$ & $-1$ & $1$ & $1$ & $1$ & $-1$ \\ \hline
		$I2_{1\bar{1}0}$ &					$\left(\begin{array}{cc} 0&i \\ i&0 \end{array}\right)$ &$-i$ & $i$ & $i$ & $-i$ & $0$ & $0$ \\ \hline
		$I2_{01\bar{1}}$ &				$\left(\begin{array}{cc} 0&-(\sqrt{3}-i)/2 \\ (\sqrt{3}+i)/2&0 \end{array}\right)$ &$i$ & $-i$ & $-i$ & $i$ & $0$ & $0$ \\ \hline
		$I2_{\bar{1}01}$ &		$\left(\begin{array}{cc} 0&-(\sqrt{3}+i)/2 \\ (\sqrt{3}-i)/2&0 \end{array}\right)$ &$-i$ & $i$ & $i$ & $-i$ & $0$ & $0$ \\ \hline
	\end{tabular}
\end{table}

\restylefloat{table}
\begin{table}[H]
	\begin{tabular}{|c|c||c|c|c|c|} \hline
		$p_g$ & $p_g^{\text{sp}}$ & $\bar{L}_3$ & $\bar{L}_4$ & $\bar{L}_5$ & $\bar{L}_6$  \\ \hline \hline
		$1$ & 					$\left(\begin{array}{cc} 1&0 \\ 0&1 \end{array}\right)$ &$1$ &  $1$ &  $1$ &  $1$  \\ \hline
		$2_{\bar{1}01}$ &		$\left(\begin{array}{cc} 0&-(\sqrt{3}+i)/2 \\ (\sqrt{3}-i)/2&0 \end{array}\right)$ &$-i$ & $i$ & $-i$ & $i$ \\ \hline
		$I$ & 					$\left(\begin{array}{cc} 1&0 \\ 0&1 \end{array}\right)$ &$1$ & $1$ & $-1$ & $-1$ \\ \hline
		$I2_{\bar{1}01}$ &		$\left(\begin{array}{cc} 0&-(\sqrt{3}+i)/2 \\ (\sqrt{3}-i)/2&0 \end{array}\right)$ &$-i$ & $i$ & $i$ & $-i$ \\ \hline
	\end{tabular}
\end{table}

\restylefloat{table}
\begin{table}[H]
	\begin{tabular}{|c|c||c|c|c|c|} \hline
		$p_g$ & $p_g^{\text{sp}}$ & $\bar{F}_3$ & $\bar{F}_4$ & $\bar{F}_5$ & $\bar{F}_6$  \\ \hline \hline
		$1$ & 							$\left(\begin{array}{cc} 1&0 \\ 0&1 \end{array}\right)$ & $1$ &  $1$ &  $1$ &  $1$  \\ \hline
		$2_{1\bar{1}0}$ &					$\left(\begin{array}{cc} 0&i \\ i&0 \end{array}\right)$ & $-i$ & $i$ & $-i$ & $i$ \\ \hline
		$I$ & 							$\left(\begin{array}{cc} 1&0 \\ 0&1 \end{array}\right)$ & $1$ & $1$ & $-1$ & $-1$ \\ \hline
		$I2_{1\bar{1}0}$ &					$\left(\begin{array}{cc} 0&i \\ i&0 \end{array}\right)$ & $-i$ & $i$ & $i$ & $-i$ \\ \hline
	\end{tabular}
\end{table}

\restylefloat{table}
\begin{table}[H]
	\begin{tabular}{|c|c||c|c|c|c|c|c|} \hline
		$p_g$ & $p_g^{\text{sp}}$ & $\bar{T}_4$ & $\bar{T}_5$ & $\bar{T}_6$ & $\bar{T}_7$ & $\bar{T}_8$ & $\bar{T}_9$  \\ \hline \hline
		$1$ & 							$\left(\begin{array}{cc} 1&0 \\ 0&1 \end{array}\right)$ & $1$ &  $1$ &  $1$ &  $1$ &  $2$ &  $2$ \\ \hline
		$3^{+}_{001}$ &				$\left(\begin{array}{cc} (1+\sqrt{3}i)/2&0 \\ 0&(1-\sqrt{3}i)/2 \end{array}\right)$ & $-1$ & $-1$ & $-1$ & $-1$ & $1$ & $1$ \\ \hline
		$3^{-}_{001}$ &				$\left(\begin{array}{cc} (1-\sqrt{3}i)/2&0 \\ 0&(1+\sqrt{3}i)/2 \end{array}\right)$ &$-1$ & $-1$ & $-1$ & $-1$ & $1$ & $1$ \\ \hline
		$2_{1\bar{1}0}$ &					$\left(\begin{array}{cc} 0&i \\ i&0 \end{array}\right)$ & $-i$ & $i$ & $-i$ & $i$ & $0$ & $0$ \\ \hline
		$2_{01\bar{1}}$ &				$\left(\begin{array}{cc} 0&-(\sqrt{3}-i)/2 \\ (\sqrt{3}+i)/2&0 \end{array}\right)$ & $i$ & $-i$ & $i$ & $-i$ & $0$ & $0$ \\ \hline
		$2_{\bar{1}01}$ &		$\left(\begin{array}{cc} 0&-(\sqrt{3}+i)/2 \\ (\sqrt{3}-i)/2&0 \end{array}\right)$ & $-i$ & $i$ & $-i$ & $i$ & $0$ & $0$ \\ \hline
		$I$ & 					$\left(\begin{array}{cc} 1&0 \\ 0&1 \end{array}\right)$ & $1$ & $1$ & $-1$ & $-1$ & $2$ & $-2$ \\ \hline
		$I3^{+}_{001}$ &		$\left(\begin{array}{cc} (1+\sqrt{3}i)/2&0 \\ 0&(1-\sqrt{3}i)/2 \end{array}\right)$ &$-1$ & $-1$ & $1$ & $1$ & $1$ & $-1$ \\ \hline
		$I3^{-}_{001}$ &		$\left(\begin{array}{cc} (1-\sqrt{3}i)/2&0 \\ 0&(1+\sqrt{3}i)/2 \end{array}\right)$ &$-1$ & $-1$ & $1$ & $1$ & $1$ & $-1$ \\ \hline
		$I2_{1\bar{1}0}$ &					$\left(\begin{array}{cc} 0&i \\ i&0 \end{array}\right)$ &$-i$ & $i$ & $i$ & $-i$ & $0$ & $0$ \\ \hline
		$I2_{01\bar{1}}$ &				$\left(\begin{array}{cc} 0&-(\sqrt{3}-i)/2 \\ (\sqrt{3}+i)/2&0 \end{array}\right)$ &$i$ & $-i$ & $-i$ & $i$ & $0$ & $0$ \\ \hline
		$I2_{\bar{1}01}$ &		$\left(\begin{array}{cc} 0&-(\sqrt{3}+i)/2 \\ (\sqrt{3}-i)/2&0 \end{array}\right)$ &$-i$ & $i$ & $i$ & $-i$ & $0$ & $0$ \\ \hline
	\end{tabular}
\end{table}

\subsection{Irreducible representations of bismuth}
\label{app:n_BI}
\restylefloat{table}
\begin{table}[H]
	\begin{tabular}{|c|c|}\hline
		$\nu$ & 30 \\ \hline \hline
		$n_{\Gamma}^{4}$ & $3$\\ \hline
		$n_{\Gamma}^{5}$ & $3$\\ \hline
		$n_{\Gamma}^{6}$ & $2$\\ \hline
		$n_{\Gamma}^{7}$ & $2$\\ \hline
		$n_{\Gamma}^{8}$ & $6$\\ \hline
		$n_{\Gamma}^{9}$ & $4$\\ \hline \hline
		$n_{L}^{3}$ & $7$\\ \hline
		$n_{L}^{4}$ & $7$\\ \hline
		$n_{L}^{5}$ & $8$\\ \hline
		$n_{L}^{6}$ & $8$\\ \hline \hline
		$n_{F}^{3}$ & $7$\\ \hline
		$n_{F}^{4}$ & $7$\\ \hline
		$n_{F}^{5}$ & $8$\\ \hline
		$n_{F}^{6}$ & $8$\\ \hline \hline
		$n_{T}^{4}$ & $2$\\ \hline
		$n_{T}^{5}$ & $2$\\ \hline
		$n_{T}^{6}$ & $3$\\ \hline
		$n_{T}^{7}$ & $3$\\ \hline
		$n_{T}^{8}$ & $5$\\ \hline 
		$n_{T}^{9}$ & $5$\\ \hline 
	\end{tabular}
\end{table}

\subsection{Character tables for silicon~\cite{Bilbao}}\label{app:ch-si}

\restylefloat{table}
\begin{table}[H]
	\scalebox{0.9}[0.9]{
		\begin{tabular}{|l|c||c|c|c|c|c|c|} \hline
			$\{p_g|\bt_g\}$ & $p_g^{\text{sp}}$ & $\bar{\Gamma}_6$ & $\bar{\Gamma}_7$ & $\bar{\Gamma}_8$ & $\bar{\Gamma}_9$ & $\bar{\Gamma}_{10}$ & $\bar{\Gamma}_{11}$  \\ \hline \hline
			$\{1|0,0,0\}$ & 				$\left(\begin{array}{cc} 1&0 \\ 0&1 \end{array}\right)$	& $2$ &  $2$ &  $2$ &  $2$ &  $4$ &  $4$ \\ \hline
			$\{2_{001}|-1/4,-1/4,0\}$ &		$\left(\begin{array}{cc} -i&0 \\ 0&i \end{array}\right)$	& $0$ & $0$ & $0$ & $0$ & $0$ & $0$ \\ \hline
			$\{2_{010}|-1/4,0,-1/4\}$ &		$\left(\begin{array}{cc} 0&-1 \\ 1&0 \end{array}\right)$	&$0$ & $0$ & $0$ & $0$ & $0$ & $0$ \\ \hline
			$\{2_{100}|0,-1/4,-1/4\}$ &		$\left(\begin{array}{cc} 0&-i \\ -i&0 \end{array}\right)$	& $0$ & $0$ & $0$ & $0$ & $0$ & $0$ \\ \hline
			$\{3^{+}_{111}|0,0,0\}$ &		$\left(\begin{array}{cc} (1-i)/2&-(1+i)/2 \\ (1-i)/2&(1+i)/2 \end{array}\right)$	&$1$ & $1$ & $1$ & $1$ & $-1$ & $-1$ \\ \hline
			$\{3^{+}_{\bar{1}1\bar{1}}|0,-1/4,-1/4\}$ &		$\left(\begin{array}{cc} (1+i)/2&-(1-i)/2 \\ (1+i)/2&(1-i)/2 \end{array}\right)$	&$1$ & $1$ & $1$ & $1$ & $-1$ & $-1$ \\ \hline
			$\{3^{+}_{1\bar{1}\bar{1}}|-1/4,-1/4,0\}$ &		$\left(\begin{array}{cc} (1+i)/2&(1-i)/2 \\ -(1+i)/2&(1-i)/2 \end{array}\right)$	& $1$ & $1$ & $1$ & $1$ & $-1$ & $-1$ \\
\hline
		\end{tabular}
		}
\end{table}
			
\restylefloat{table}
\begin{table}[H]
	\scalebox{0.9}[0.9]{
		\begin{tabular}{|l|c||c|c|c|c|c|c|} \hline									
			$\{I|0,0,0\}$ & 				$\left(\begin{array}{cc} 1&0 \\ 0&1 \end{array}\right)$	& 2 &  2 &  -2 &  -2 &  4 &  -4 \\ \hline
			$\{I2_{001}|1/4,1/4,0\}$ &		$\left(\begin{array}{cc} -i&0 \\ 0&i \end{array}\right)$	& 0 & $0$ & 0 & $0$ & 0 & 0 \\ \hline
			$\{I2_{010}|1/4,0,1/4\}$ &		$\left(\begin{array}{cc} 0&-1 \\ 1&0 \end{array}\right)$	&0 & $0$ & 0 & $0$ & 0 & 0 \\ \hline
			$\{I2_{100}|0,1/4,1/4\}$ &		$\left(\begin{array}{cc} 0&-i \\ -i&0 \end{array}\right)$	& 0 & $0$ & 0 & $0$ & 0 & 0 \\ \hline
			$\{I3^{+}_{111}|0,0,0\}$ &		$\left(\begin{array}{cc} (1-i)/2&-(1+i)/2 \\ (1-i)/2&(1+i)/2 \end{array}\right)$	&1 & 1 & $-1$ & $-1$ & $-1$ & 1 \\ \hline
			$\{I3^{+}_{\bar{1}1\bar{1}}|0,1/4,1/4\}$ &		$\left(\begin{array}{cc} (1+i)/2&-(1-i)/2 \\ (1+i)/2&(1-i)/2 \end{array}\right)$	&1 & 1 & $-1$ & $-1$ & $-1$ & 1 \\ \hline
			$\{I3^{+}_{1\bar{1}\bar{1}}|1/4,1/4,0\}$ &		$\left(\begin{array}{cc} (1+i)/2&(1-i)/2 \\ -(1+i)/2&(1-i)/2 \end{array}\right)$	& 1 & 1 & $-1$ & $-1$ & $-1$ & 1 \\ \hline
			$\{I3^{+}_{\bar{1}\bar{1}1}|1/4,0,1/4\}$ &		$\left(\begin{array}{cc} (1-i)/2&(1+i)/2 \\ -(1-i)/2&(1+i)/2 \end{array}\right)$	&1 & 1 & $-1$ & $-1$ & $-1$ & 1 \\
\hline
		\end{tabular}
		}
\end{table}

\restylefloat{table}
\begin{table}[H]
	\begin{tabular}{|l|c||c|c|c|c|c|c|} \hline
		$\{p_g|\bt_g\}$ & $p_g^{\text{sp}}$ & $\bar{L}_4$ & $\bar{L}_5$ & $\bar{L}_6$ & $\bar{L}_7$ & $\bar{L}_{8}$ & $\bar{L}_{9}$  \\ \hline \hline
		$\{1|0,0,0\}$ & 				$\left(\begin{array}{cc} 1&0 \\ 0&1 \end{array}\right)$ & $1$ &  $1$ &  $1$ &  $1$ &  $2$ &  $2$ \\ \hline
		$\{3^{+}_{111}|0,0,0\}$ &		$\left(\begin{array}{cc} (1-i)/2&-(1+i)/2 \\ (1-i)/2&(1+i)/2 \end{array}\right)$	&$-1$ & $-1$ & $-1$ & $-1$ & $1$ & $1$ \\ \hline
		$\{3^{-}_{111}|0,0,0\}$ &		$\left(\begin{array}{cc} (1+i)/2&(1+i)/2 \\ -(1-i)/2&(1-i)/2 \end{array}\right)$	&$-1$ & $-1$ & $-1$ & $-1$ & $1$ & $1$ \\ \hline
		$\{2_{1\bar{1}0}|0,0,0\}$ &	$\left(\begin{array}{cc} 0&-(1-i)\sqrt{2}/2 \\ (1+i)\sqrt{2}/2&0 \end{array}\right)$	&$-i$ & $i$ & $-i$ & $i$ & $0$ & $0$ \\ \hline
		$\{2_{01\bar{1}}|0,0,0\}$ &	$\left(\begin{array}{cc} -i\sqrt{2}/2&\sqrt{2}/2 \\ -\sqrt{2}/2&i\sqrt{2}/2 \end{array}\right)$	&$-i$ & $i$ & $-i$ & $i$ & $0$ & $0$ \\ \hline
		$\{2_{\bar{1}01}|0,0,0\}$ &	$\left(\begin{array}{cc} -i\sqrt{2}/2&i\sqrt{2}/2 \\ i\sqrt{2}/2&i\sqrt{2}/2 \end{array}\right)$	&$i$ & $-i$ & $i$ & $-i$ & $0$ & $0$ \\ \hline
		$\{I|0,0,0\}$ & 				 	$\left(\begin{array}{cc} 1&0 \\ 0&1 \end{array}\right)$ &$1$ &  $1$ &  $-1$ &  $-1$ &  $2$ &  $-2$ \\
\hline
	\end{tabular}
\end{table}

\restylefloat{table}
\begin{table}[H]
	\begin{tabular}{|l|c||c|}\hline
		$\{p_g|\bt_g\}$ & $p_g^{\text{sp}}$ & $\bar{X}_5$ \\ \hline \hline
		$\{1|0,0,0\}$ & 				$\left(\begin{array}{cc} 1&0 \\ 0&1 \end{array}\right)$	& $4$  \\ \hline
		$\{2_{001}|-1/4,-1/4,0\}$ &		$\left(\begin{array}{cc} -i&0 \\ 0&i \end{array}\right)$	& $0$  \\ \hline
		$\{2_{010}|-1/4,0,-1/4\}$ &		$\left(\begin{array}{cc} 0&-1 \\ 1&0 \end{array}\right)$	& $0$  \\ \hline
		$\{2_{100}|0,-1/4,-1/4\}$ &		$\left(\begin{array}{cc} 0&-i \\ -i&0 \end{array}\right)$	& $0$  \\ \hline
		$\{4^{+}_{010}|1/4,1/4,0\}$ &	$\left(\begin{array}{cc} \sqrt{2}/2&-\sqrt{2}/2 \\ \sqrt{2}/2&\sqrt{2}/2 \end{array}\right)$	&$0$ \\ \hline
		$\{2_{101}|1/4,0,1/4\}$ &		$\left(\begin{array}{cc} -i\sqrt{2}/2&-i\sqrt{2}/2 \\ -i\sqrt{2}/2&i\sqrt{2}/2 \end{array}\right)$	&$0$ \\ \hline
		$\{4^{-}_{010}|0,1/4,1/4\}$ &		$\left(\begin{array}{cc} \sqrt{2}/2&\sqrt{2}/2 \\ -\sqrt{2}/2&\sqrt{2}/2 \end{array}\right)$	&$0$ \\
\hline
	\end{tabular}
\end{table}

\restylefloat{table}
\begin{table}[H]
	\begin{tabular}{|l|c||c|c|c|c|c|} \hline
		$\{p_g|\bt_g\}$ & $p_g^{\text{sp}}$ & $\bar{W}_3$ & $\bar{W}_4$ & $\bar{W}_5$ & $\bar{W}_6$ & $\bar{W}_{7}$  \\ \hline \hline
		$\{1|0,0,0\}$ & 				$\left(\begin{array}{cc} 1&0 \\ 0&1 \end{array}\right)$ &$1$ &  $1$ &  $1$ &  $1$ &  $2$ \\ \hline
		$\{2_{100}|0,-1/4,-1/4\}$ &		$\left(\begin{array}{cc} 0&-i \\ -i&0 \end{array}\right)$	& $i$ & $i$ & $i$ & $i$ & $-2i$ \\ \hline
		$\{2_{011}|0,1/4,1/4\}$ &		$\left(\begin{array}{cc} -i\sqrt{2}/2&-\sqrt{2}/2 \\ \sqrt{2}/2&i\sqrt{2}/2 \end{array}\right)$	&$-1$ & $1$ & $-1$ & $1$ & $0$ \\ \hline
		$\{2_{01\bar{1}}|0,0,0\}$ &		$\left(\begin{array}{cc} -i\sqrt{2}/2&\sqrt{2}/2 \\ -\sqrt{2}/2&i\sqrt{2}/2 \end{array}\right)$	&$i$ & $-i$ & $i$ & $-i$ & $0$ \\ \hline
		$\{I2_{001}|1/4,1/4,0\}$ &		$\left(\begin{array}{cc} -i&0 \\ 0&i \end{array}\right)$	&$e^{-i\pi/4}$ & $e^{i 3\pi/4}$ & $e^{i 3\pi/4}$ & $e^{-i\pi/4}$ & 0 \\ \hline
		$\{I2_{010}|1/4,0,1/4\}$ &		$\left(\begin{array}{cc} 0&-1 \\ 1&0 \end{array}\right)$	&$e^{i\pi/4}$ & $e^{-i 3\pi/4}$ & $e^{-i 3\pi/4}$ & $e^{i\pi/4}$ & 0 \\ \hline
		$\{I4^{-}_{100}|-1/4,-1/4,0\}$ &		$\left(\begin{array}{cc} \sqrt{2}/2&i\sqrt{2}/2 \\ i\sqrt{2}/2&\sqrt{2}/2 \end{array}\right)$	&$e^{i3\pi/4}$ & $e^{i 3\pi/4}$ & $e^{-i \pi/4}$ & $e^{-i\pi/4}$ & 0 \\ \hline
		$\{I4^{+}_{100}|-1/4,0,-1/4\}$ &		$\left(\begin{array}{cc} \sqrt{2}/2&-i\sqrt{2}/2 \\ -i\sqrt{2}/2&\sqrt{2}/2 \end{array}\right)$	&$e^{-i3\pi/4}$ & $e^{-i 3\pi/4}$ & $e^{i \pi/4}$ & $e^{i\pi/4}$ & 0 \\ \hline
	\end{tabular}
\end{table}


\subsection{Irreducible representations of silicon}\label{app:b-si}

\restylefloat{table}
\begin{table}[H]
	\begin{tabular}{|c|c|}\hline
		$\nu$ & 8 \\ \hline \hline
		$n_{\Gamma}^{6}$ & $1$\\ \hline
		$n_{\Gamma}^{7}$ & $1$\\ \hline
		$n_{\Gamma}^{8}$ & $0$\\ \hline
		$n_{\Gamma}^{9}$ & $0$\\ \hline
		$n_{\Gamma}^{{10}}$ & $1$\\ \hline
		$n_{\Gamma}^{{11}}$ & $0$\\ \hline \hline
		$n_{L}^{4}$ & $1$\\ \hline
		$n_{L}^{5}$ & $1$\\ \hline
		$n_{L}^{6}$ & $0$\\ \hline
		$n_{L}^{7}$ & $0$\\ \hline
		$n_{L}^{8}$ & $2$\\ \hline
		$n_{L}^{9}$ & $1$\\ \hline \hline
		$n_{W}^{3}$ & $1$\\ \hline
		$n_{W}^{4}$ & $1$\\ \hline
		$n_{W}^{5}$ & $1$\\ \hline
		$n_{W}^{6}$ & $1$\\ \hline
		$n_{W}^{7}$ & $2$\\ \hline 
	\end{tabular}
\end{table}

\subsection{Character tables for NaCdAs~\cite{Bilbao}}\label{app:ch-NaCdAs}
\restylefloat{table}
\begin{table}[H]
	\begin{tabular}{|l|c||c|c|}\hline
			$\{p_g|\bt_g\}$ & $p_g^{\text{sp}}$ & $\bar{\Gamma}_5$ & $\bar{\Gamma}_6$  \\ \hline \hline
			$\{1|0,0,0\}$ & 				$\left(\begin{array}{cc} 1&0 \\ 0&1 \end{array}\right)$	& $2$ &  $2$ \\ \hline
			$\{2_{001}|-1/2,0,1/2\}$ &	$\left(\begin{array}{cc} -i&0 \\ 0&i \end{array}\right)$	& $0$ & $0$  \\ \hline
			$\{2_{010}|0,1/2,0\}$ &		$\left(\begin{array}{cc} 0&-1 \\ 1&0 \end{array}\right)$	& $0$ & $0$\\ \hline
			$\{2_{100}|1/2,-1/2,-1/2\}$ &	$\left(\begin{array}{cc} 0&-i \\ -i&0 \end{array}\right)$	& $0$ & $0$\\ \hline
			$\{I|0,0,0\}$ & 					$\left(\begin{array}{cc} 1&0 \\ 0&1 \end{array}\right)$	& $2$ &  $-2$\\ \hline
			$\{I2_{001}|1/2,0,-1/2\}$ &	$\left(\begin{array}{cc} -i&0 \\ 0&i \end{array}\right)$	& $0$ & $0$  \\ \hline
			$\{I2_{010}|0,-1/2,0\}$ &		$\left(\begin{array}{cc} 0&-1 \\ -1&0 \end{array}\right)$	& $0$ & $0$\\ \hline
			$\{I2_{100}|-1/2,1/2,1/2\}$ &	$\left(\begin{array}{cc} 0&-i \\ -i&0 \end{array}\right)$	& $0$ & $0$\\ \hline
	\end{tabular}
\end{table}

\restylefloat{table}
\begin{table}[H]
		\begin{tabular}{|l|c||c|c|}\hline
			$\{p_g|\bt_g\}$ & $p_g^{\text{sp}}$ & $\bar{R}_3$ & $\bar{R}_4$  \\ \hline \hline
			$\{1|0,0,0\}$ & 					$\left(\begin{array}{cc} 1&0 \\ 0&1 \end{array}\right)$	& $2$ &  $2$ \\ \hline
			$\{2_{001}|-1/2,0,1/2\}$ &		$\left(\begin{array}{cc} -i&0 \\ 0&i \end{array}\right)$	& $2$ & $-2$  \\ \hline
			$\{2_{010}|0,1/2,0\}$ &			$\left(\begin{array}{cc} 0&-1 \\ 1&0 \end{array}\right)$	& $0$ & $0$\\ \hline
			$\{2_{100}|1/2,-1/2,-1/2\}$ &		$\left(\begin{array}{cc} 0&-i \\ -i&0 \end{array}\right)$	& $0$ & $0$\\ \hline
			$\{I|0,0,0\}$ & 						$\left(\begin{array}{cc} 1&0 \\ 0&1 \end{array}\right)$	& $0$ &  $0$\\ \hline
			$\{I2_{001}|1/2,0,-1/2\}$ &		$\left(\begin{array}{cc} -i&0 \\ 0&i \end{array}\right)$	& $0$ & $0$  \\ \hline
			$\{I2_{010}|0,-1/2,0\}$ &			$\left(\begin{array}{cc} 0&-1 \\ -1&0 \end{array}\right)$	& $0$ & $0$\\ \hline
			$\{I2_{100}|-1/2,1/2,1/2\}$ &		$\left(\begin{array}{cc} 0&-i \\ -i&0 \end{array}\right)$	& $0$ & $0$\\ \hline
	\end{tabular}
\end{table}

\restylefloat{table}
\begin{table}[H]
		\begin{tabular}{|l|c||c|c|}\hline
			$\{p_g|\bt_g\}$ & $p_g^{\text{sp}}$ & $\bar{S}_3$ & $\bar{S}_4$  \\ \hline \hline
			$\{1|0,0,0\}$ & 					$\left(\begin{array}{cc} 1&0 \\ 0&1 \end{array}\right)$	& $2$ &  $2$ \\ \hline
			$\{2_{001}|-1/2,0,1/2\}$ &		$\left(\begin{array}{cc} -i&0 \\ 0&i \end{array}\right)$	& $0$ & $0$  \\ \hline
			$\{2_{010}|0,1/2,0\}$ &			$\left(\begin{array}{cc} 0&-1 \\ 1&0 \end{array}\right)$	& $0$ & $0$\\ \hline
			$\{2_{100}|1/2,-1/2,-1/2\}$ &		$\left(\begin{array}{cc} 0&-i \\ -i&0 \end{array}\right)$	& $0$ & $0$\\ \hline
			$\{I|0,0,0\}$ & 						$\left(\begin{array}{cc} 1&0 \\ 0&1 \end{array}\right)$	& $0$ &  $0$\\ \hline
			$\{I2_{001}|1/2,0,-1/2\}$ &		$\left(\begin{array}{cc} -i&0 \\ 0&i \end{array}\right)$	& $0$ & $0$  \\ \hline
			$\{I2_{010}|0,-1/2,0\}$ &			$\left(\begin{array}{cc} 0&-1 \\ -1&0 \end{array}\right)$	& $0$ & $0$\\ \hline
			$\{I2_{100}|-1/2,1/2,1/2\}$ &		$\left(\begin{array}{cc} 0&-i \\ -i&0 \end{array}\right)$	& $-2$ & $2$\\ \hline
	\end{tabular}
\end{table}

\restylefloat{table}
\begin{table}[H]
	\begin{tabular}{|l|c||c|c|}\hline
			$\{p_g|\bt_g\}$ & $p_g^{\text{sp}}$ & $\bar{T}_3$ & $\bar{T}_4$  \\ \hline \hline
			$\{1|0,0,0\}$ & 					$\left(\begin{array}{cc} 1&0 \\ 0&1 \end{array}\right)$	& $2$ &  $2$ \\ \hline
			$\{2_{001}|-1/2,0,1/2\}$ &		$\left(\begin{array}{cc} -i&0 \\ 0&i \end{array}\right)$	& $0$ & $0$  \\ \hline
			$\{2_{010}|0,1/2,0\}$ &			$\left(\begin{array}{cc} 0&-1 \\ 1&0 \end{array}\right)$	& $0$ & $0$\\ \hline
			$\{2_{100}|1/2,-1/2,-1/2\}$ &		$\left(\begin{array}{cc} 0&-i \\ -i&0 \end{array}\right)$	& $0$ & $0$\\ \hline
			$\{I|0,0,0\}$ & 						$\left(\begin{array}{cc} 1&0 \\ 0&1 \end{array}\right)$	& $0$ &  $0$\\ \hline
			$\{I2_{001}|1/2,0,-1/2\}$ &		$\left(\begin{array}{cc} -i&0 \\ 0&i \end{array}\right)$	& $0$ & $0$  \\ \hline
			$\{I2_{010}|0,-1/2,0\}$ &			$\left(\begin{array}{cc} 0&-1 \\ -1&0 \end{array}\right)$	& $0$ & $0$\\ \hline
			$\{I2_{100}|-1/2,1/2,1/2\}$ &		$\left(\begin{array}{cc} 0&-i \\ -i&0 \end{array}\right)$	& $-2i$ & $2i$\\ \hline
	\end{tabular}
\end{table}

\restylefloat{table}
\begin{table}[H]
	\begin{tabular}{|l|c||c|c|}\hline
			$\{p_g|\bt_g\}$ & $p_g^{\text{sp}}$ & $\bar{U}_5$ & $\bar{U}_6$  \\ \hline \hline
			$\{1|0,0,0\}$ & 					$\left(\begin{array}{cc} 1&0 \\ 0&1 \end{array}\right)$	& $2$ &  $2$ \\ \hline
			$\{2_{001}|-1/2,0,1/2\}$ &		$\left(\begin{array}{cc} -i&0 \\ 0&i \end{array}\right)$	& $0$ & $0$  \\ \hline
			$\{2_{010}|0,1/2,0\}$ &			$\left(\begin{array}{cc} 0&-1 \\ 1&0 \end{array}\right)$	& $0$ & $0$\\ \hline
			$\{2_{100}|1/2,-1/2,-1/2\}$ &		$\left(\begin{array}{cc} 0&-i \\ -i&0 \end{array}\right)$	& $0$ & $0$\\ \hline
			$\{I|0,0,0\}$ & 						$\left(\begin{array}{cc} 1&0 \\ 0&1 \end{array}\right)$	& $2$ &  $-2$\\ \hline
			$\{I2_{001}|1/2,0,-1/2\}$ &		$\left(\begin{array}{cc} -i&0 \\ 0&i \end{array}\right)$	& $0$ & $0$  \\ \hline
			$\{I2_{010}|0,-1/2,0\}$ &			$\left(\begin{array}{cc} 0&-1 \\ -1&0 \end{array}\right)$	& $0$ & $0$\\ \hline
			$\{I2_{100}|-1/2,1/2,1/2\}$ &		$\left(\begin{array}{cc} 0&-i \\ -i&0 \end{array}\right)$	& $0$ & $0$\\ \hline
	\end{tabular}
\end{table}

\restylefloat{table}
\begin{table}[H]
	\begin{tabular}{|l|c||c|c|}\hline
			$\{p_g|\bt_g\}$ & $p_g^{\text{sp}}$ & $\bar{X}_3$ & $\bar{X}_4$  \\ \hline \hline
			$\{1|0,0,0\}$ & 					$\left(\begin{array}{cc} 1&0 \\ 0&1 \end{array}\right)$	& $2$ &  $2$ \\ \hline
			$\{2_{001}|-1/2,0,1/2\}$ &		$\left(\begin{array}{cc} -i&0 \\ 0&i \end{array}\right)$	& $0$ & $0$  \\ \hline
			$\{2_{010}|0,1/2,0\}$ &			$\left(\begin{array}{cc} 0&-1 \\ 1&0 \end{array}\right)$	& $-2i$ & $2i$\\ \hline
			$\{2_{100}|1/2,-1/2,-1/2\}$ &		$\left(\begin{array}{cc} 0&-i \\ -i&0 \end{array}\right)$	& $0$ & $0$\\ \hline
			$\{I|0,0,0\}$ & 						$\left(\begin{array}{cc} 1&0 \\ 0&1 \end{array}\right)$	& $0$ &  $0$\\ \hline
			$\{I2_{001}|1/2,0,-1/2\}$ &		$\left(\begin{array}{cc} -i&0 \\ 0&i \end{array}\right)$	& $0$ & $0$  \\ \hline
			$\{I2_{010}|0,-1/2,0\}$ &			$\left(\begin{array}{cc} 0&-1 \\ -1&0 \end{array}\right)$	& $0$ & $0$\\ \hline
			$\{I2_{100}|-1/2,1/2,1/2\}$ &		$\left(\begin{array}{cc} 0&-i \\ -i&0 \end{array}\right)$	& $0$ & $0$\\ \hline
	\end{tabular}
\end{table}

\restylefloat{table}
\begin{table}[H]
	\begin{tabular}{|l|c||c|c|}\hline
			$\{p_g|\bt_g\}$ & $p_g^{\text{sp}}$ & $\bar{Y}_3$ & $\bar{Y}_4$  \\ \hline \hline
			$\{1|0,0,0\}$ & 					$\left(\begin{array}{cc} 1&0 \\ 0&1 \end{array}\right)$	& $2$ &  $2$ \\ \hline
			$\{2_{001}|-1/2,0,1/2\}$ &		$\left(\begin{array}{cc} -i&0 \\ 0&i \end{array}\right)$	& $-2i$ & $2i$  \\ \hline
			$\{2_{010}|0,1/2,0\}$ &			$\left(\begin{array}{cc} 0&-1 \\ 1&0 \end{array}\right)$	& $0$ & $0$\\ \hline
			$\{2_{100}|1/2,-1/2,-1/2\}$ &		$\left(\begin{array}{cc} 0&-i \\ -i&0 \end{array}\right)$	& $0$ & $0$\\ \hline
			$\{I|0,0,0\}$ & 						$\left(\begin{array}{cc} 1&0 \\ 0&1 \end{array}\right)$	& $0$ &  $0$\\ \hline
			$\{I2_{001}|1/2,0,-1/2\}$ &		$\left(\begin{array}{cc} -i&0 \\ 0&i \end{array}\right)$	& $0$ & $0$  \\ \hline
			$\{I2_{010}|0,-1/2,0\}$ &			$\left(\begin{array}{cc} 0&-1 \\ -1&0 \end{array}\right)$	& $0$ & $0$\\ \hline
			$\{I2_{100}|-1/2,1/2,1/2\}$ &		$\left(\begin{array}{cc} 0&-i \\ -i&0 \end{array}\right)$	& $0$ & $0$\\ \hline
	\end{tabular}
\end{table}

\restylefloat{table}
\begin{table}[H]
	\begin{tabular}{|l|c||c|c|}\hline
			$\{p_g|\bt_g\}$ & $p_g^{\text{sp}}$ & $\bar{Z}_3$ & $\bar{Z}_4$  \\ \hline \hline
			$\{1|0,0,0\}$ & 					$\left(\begin{array}{cc} 1&0 \\ 0&1 \end{array}\right)$	& $2$ &  $2$ \\ \hline
			$\{2_{001}|-1/2,0,1/2\}$ &		$\left(\begin{array}{cc} -i&0 \\ 0&i \end{array}\right)$	& $0$ & $0$  \\ \hline
			$\{2_{010}|0,1/2,0\}$ &			$\left(\begin{array}{cc} 0&-1 \\ 1&0 \end{array}\right)$	& $2i$ & $-2i$\\ \hline
			$\{2_{100}|1/2,-1/2,-1/2\}$ &		$\left(\begin{array}{cc} 0&-i \\ -i&0 \end{array}\right)$	& $0$ & $0$\\ \hline
			$\{I|0,0,0\}$ & 						$\left(\begin{array}{cc} 1&0 \\ 0&1 \end{array}\right)$	& $0$ &  $0$\\ \hline
			$\{I2_{001}|1/2,0,-1/2\}$ &		$\left(\begin{array}{cc} -i&0 \\ 0&i \end{array}\right)$	& $0$ & $0$  \\ \hline
			$\{I2_{010}|0,-1/2,0\}$ &			$\left(\begin{array}{cc} 0&-1 \\ -1&0 \end{array}\right)$	&  $0$ & $0$\\ \hline
			$\{I2_{100}|-1/2,1/2,1/2\}$ &		$\left(\begin{array}{cc} 0&-i \\ -i&0 \end{array}\right)$	& $0$ & $0$\\ \hline
	\end{tabular}
\end{table}

\subsection{Irreducible representations of NaCdAs}\label{app:b-NaCdAs}

\restylefloat{table}
\begin{table}[H]
	\begin{tabular}{|c|c|}\hline
		$\nu$ & 176 \\ \hline \hline
		$n_{\Gamma}^{5}$ & $45$\\ \hline
		$n_{\Gamma}^{6}$ & $43$\\ \hline \hline
		$n_{R}^{3}$ & $44$\\ \hline
		$n_{R}^{4}$ & $44$\\ \hline \hline
		$n_{S}^{3}$ & $44$\\ \hline 
		$n_{S}^{4}$ & $44$\\ \hline  \hline
		$n_{T}^{3}$ & $44$\\ \hline
		$n_{T}^{4}$ & $44$\\ \hline \hline
		$n_{U}^{5}$ & $44$\\ \hline
		$n_{U}^{6}$ & $44$\\ \hline \hline
		$n_{X}^{3}$ & $44$\\ \hline
		$n_{X}^{4}$ & $44$\\ \hline \hline
		$n_{Y}^{3}$ & $44$\\ \hline
		$n_{Y}^{4}$ & $44$\\ \hline \hline
		$n_{Z}^{3}$ & $44$\\ \hline
		$n_{Z}^{4}$ & $44$\\ \hline
	\end{tabular}
\end{table}

\subsection{Character table for $PbPt_3$~\cite{Bilbao}}\label{app:ch-pbpt3}

\restylefloat{table}
\begin{table}[H]
	\scalebox{0.9}[0.9]{
		\begin{tabular}{|c|c||c|c|c|c|c|c|}\hline
			$p_g$ & $p_g^{\text{sp}}$ & $\bar{\Gamma}_6$ & $\bar{\Gamma}_7$ & $\bar{\Gamma}_8$ & $\bar{\Gamma}_9$ & $\bar{\Gamma}_{10}$ & $\bar{\Gamma}_{11}$  \\ \hline \hline
			$1$ & 				$\left(\begin{array}{cc} 1&0 \\ 0&1 \end{array}\right)$	& $2$ &  $2$ &  $2$ &  $2$ &  $4$ &  $4$ \\ \hline
			$2_{001}$ &		$\left(\begin{array}{cc} -i&0 \\ 0&i \end{array}\right)$	& $0$ & $0$ & $0$ & $0$ & $0$ & $0$ \\ \hline
			$2_{010}$ &		$\left(\begin{array}{cc} 0&-1 \\ 1&0 \end{array}\right)$	&$0$ & $0$ & $0$ & $0$ & $0$ & $0$ \\ \hline
			$2_{100}$ &		$\left(\begin{array}{cc} 0&-i \\ -i&0 \end{array}\right)$	& $0$ & $0$ & $0$ & $0$ & $0$ & $0$ \\ \hline
			$3^{+}_{111}$ &		$\left(\begin{array}{cc} (1-i)/2&-(1+i)/2 \\ (1-i)/2&(1+i)/2 \end{array}\right)$	&$1$ & $1$ & $1$ & $1$ & $-1$ & $-1$ \\ \hline
			$3^{+}_{\bar{1}1\bar{1}}$ &		$\left(\begin{array}{cc} (1+i)/2&-(1-i)/2 \\ (1+i)/2&(1-i)/2 \end{array}\right)$	&$1$ & $1$ & $1$ & $1$ & $-1$ & $-1$ \\ \hline
			$3^{+}_{1\bar{1}\bar{1}}$ &		$\left(\begin{array}{cc} (1+i)/2&(1-i)/2 \\ -(1+i)/2&(1-i)/2 \end{array}\right)$	& $1$ & $1$ & $1$ & $1$ & $-1$ & $-1$ \\
\hline
		\end{tabular}
	}
\end{table}

\restylefloat{table}
\begin{table}[H]
	\scalebox{0.9}[0.9]{
		\begin{tabular}{|c|c||c|c|c|c|c|c|} \hline									
			$I$ & 				$\left(\begin{array}{cc} 1&0 \\ 0&1 \end{array}\right)$	& 2 &  2 &  -2 &  -2 &  4 &  -4 \\ \hline
			$I2_{001}$ &		$\left(\begin{array}{cc} -i&0 \\ 0&i \end{array}\right)$	& 0 & $0$ & 0 & $0$ & 0 & 0 \\ \hline
			$I2_{010}$ &		$\left(\begin{array}{cc} 0&-1 \\ 1&0 \end{array}\right)$	&0 & $0$ & 0 & $0$ & 0 & 0 \\ \hline
			$I2_{100}$ &		$\left(\begin{array}{cc} 0&-i \\ -i&0 \end{array}\right)$	& 0 & $0$ & 0 & $0$ & 0 & 0 \\ \hline
			$I3^{+}_{111}$ &		$\left(\begin{array}{cc} (1-i)/2&-(1+i)/2 \\ (1-i)/2&(1+i)/2 \end{array}\right)$	&1 & 1 & $-1$ & $-1$ & $-1$ & 1 \\ \hline
			$I3^{+}_{\bar{1}1\bar{1}}$ &		$\left(\begin{array}{cc} (1+i)/2&-(1-i)/2 \\ (1+i)/2&(1-i)/2 \end{array}\right)$	&1 & 1 & $-1$ & $-1$ & $-1$ & 1 \\ \hline
			$I3^{+}_{1\bar{1}\bar{1}}$ &		$\left(\begin{array}{cc} (1+i)/2&(1-i)/2 \\ -(1+i)/2&(1-i)/2 \end{array}\right)$	& 1 & 1 & $-1$ & $-1$ & $-1$ & 1 \\ \hline
			$I3^{+}_{\bar{1}\bar{1}1}$ &		$\left(\begin{array}{cc} (1-i)/2&(1+i)/2 \\ -(1-i)/2&(1+i)/2 \end{array}\right)$	&1 & 1 & $-1$ & $-1$ & $-1$ & 1 \\
\hline
		\end{tabular}
		}
\end{table}

\restylefloat{table}
\begin{table}[H]
	\begin{tabular}{|c|c||c|c|c|c|}\hline
		$p_g$ & $p_g^{\text{sp}}$ & $\bar{M}_6$ & $\bar{M}_7$ & $\bar{M}_8$ & $\bar{M}_9$   \\ \hline \hline
		$1$ & 				$\left(\begin{array}{cc} 1&0 \\ 0&1 \end{array}\right)$	& $2$ &  $2$ &  $2$ &  $2$ \\ \hline
		$2_{001}$ &		$\left(\begin{array}{cc} -i&0 \\ 0&i \end{array}\right)$	& $0$ & $0$ & $0$ & $0$  \\ \hline
		$2_{010}$ &		$\left(\begin{array}{cc} 0&-1 \\ 1&0 \end{array}\right)$	& $0$ & $0$ & $0$ & $0$  \\ \hline
		$2_{100}$ &		$\left(\begin{array}{cc} 0&-i \\ -i&0 \end{array}\right)$	& $0$ & $0$ & $0$ & $0$  \\ \hline
		$2_{110}$ &		$\left(\begin{array}{cc} 0&-(1+i)\sqrt{2}/2 \\ (1-i)\sqrt{2}/2&0 \end{array}\right)$	&$0$ & $0$ & $0$ & $0$  \\ \hline
		$2_{1\bar{1}0}$ &		$\left(\begin{array}{cc} 0&-(1-i)\sqrt{2}/2 \\ (1+i)\sqrt{2}/2&0 \end{array}\right)$	&$0$ & $0$ & $0$ & $0$  \\ \hline
		$4^{-}_{001}$ &		$\left(\begin{array}{cc} (1+i)\sqrt{2}/2&0 \\ 0&(1-i)\sqrt{2}/2 \end{array}\right)$	&$-\sqrt{2}$ & $\sqrt{2}$ & $-\sqrt{2}$ & $\sqrt{2}$  \\
\hline
		
	\end{tabular}
\end{table}

\restylefloat{table}
\begin{table}[H]
	\begin{tabular}{|c|c||c|c|c|c|c|c|}\hline
		$p_g$ & $p_g^{\text{sp}}$ & $\bar{R}_6$ & $\bar{R}_7$ & $\bar{R}_8$ & $\bar{R}_9$ & $\bar{R}_{10}$ & $\bar{R}_{11}$  \\ \hline \hline
			$1$ & 				$\left(\begin{array}{cc} 1&0 \\ 0&1 \end{array}\right)$	& $2$ &  $2$ &  $2$ &  $2$ &  $4$ &  $4$ \\ \hline
			$2_{001}$ &		$\left(\begin{array}{cc} -i&0 \\ 0&i \end{array}\right)$	& $0$ & $0$ & $0$ & $0$ & $0$ & $0$ \\ \hline
			$2_{010}$ &		$\left(\begin{array}{cc} 0&-1 \\ 1&0 \end{array}\right)$	&$0$ & $0$ & $0$ & $0$ & $0$ & $0$ \\ \hline
			$2_{100}$ &		$\left(\begin{array}{cc} 0&-i \\ -i&0 \end{array}\right)$	& $0$ & $0$ & $0$ & $0$ & $0$ & $0$ \\ \hline
			$3^{+}_{111}$ &		$\left(\begin{array}{cc} (1-i)/2&-(1+i)/2 \\ (1-i)/2&(1+i)/2 \end{array}\right)$	&$1$ & $1$ & $1$ & $1$ & $-1$ & $-1$ \\ \hline
			$3^{+}_{\bar{1}1\bar{1}}$ &		$\left(\begin{array}{cc} (1+i)/2&-(1-i)/2 \\ (1+i)/2&(1-i)/2 \end{array}\right)$	&$1$ & $1$ & $1$ & $1$ & $-1$ & $-1$ \\ \hline
			$3^{+}_{1\bar{1}\bar{1}}$ &		$\left(\begin{array}{cc} (1+i)/2&(1-i)/2 \\ -(1+i)/2&(1-i)/2 \end{array}\right)$	& $1$ & $1$ & $1$ & $1$ & $-1$ & $-1$ \\
\hline
	\end{tabular}
\end{table}

\restylefloat{table}
\begin{table}[H]
	\begin{tabular}{|c|c||c|c|c|c|c|c|} \hline									
			$I$ & 				$\left(\begin{array}{cc} 1&0 \\ 0&1 \end{array}\right)$	& 2 &  2 &  -2 &  -2 &  4 &  -4 \\ \hline
			$I2_{001}$ &		$\left(\begin{array}{cc} -i&0 \\ 0&i \end{array}\right)$	& 0 & $0$ & 0 & $0$ & 0 & 0 \\ \hline
			$I2_{010}$ &		$\left(\begin{array}{cc} 0&-1 \\ 1&0 \end{array}\right)$	&0 & $0$ & 0 & $0$ & 0 & 0 \\ \hline
			$I2_{100}$ &		$\left(\begin{array}{cc} 0&-i \\ -i&0 \end{array}\right)$	& 0 & $0$ & 0 & $0$ & 0 & 0 \\ \hline
			$I3^{+}_{111}$ &		$\left(\begin{array}{cc} (1-i)/2&-(1+i)/2 \\ (1-i)/2&(1+i)/2 \end{array}\right)$	&1 & 1 & $-1$ & $-1$ & $-1$ & 1 \\ \hline
			$I3^{+}_{\bar{1}1\bar{1}}$ &		$\left(\begin{array}{cc} (1+i)/2&-(1-i)/2 \\ (1+i)/2&(1-i)/2 \end{array}\right)$	&1 & 1 & $-1$ & $-1$ & $-1$ & 1 \\ \hline
			$I3^{+}_{1\bar{1}\bar{1}}$ &		$\left(\begin{array}{cc} (1+i)/2&(1-i)/2 \\ -(1+i)/2&(1-i)/2 \end{array}\right)$	& 1 & 1 & $-1$ & $-1$ & $-1$ & 1 \\ \hline
			$I3^{+}_{\bar{1}\bar{1}1}$ &		$\left(\begin{array}{cc} (1-i)/2&(1+i)/2 \\ -(1-i)/2&(1+i)/2 \end{array}\right)$	&1 & 1 & $-1$ & $-1$ & $-1$ & 1 \\
\hline
	\end{tabular}
\end{table}

\restylefloat{table}
\begin{table}[H]
	\begin{tabular}{|c|c||c|c|c|c|}\hline
		$p_g$ & $p_g^{\text{sp}}$ & $\bar{X}_6$ & $\bar{X}_7$ & $\bar{X}_8$ & $\bar{X}_9$   \\ \hline \hline
		$1$ & 				$\left(\begin{array}{cc} 1&0 \\ 0&1 \end{array}\right)$	& $2$ &  $2$ & $2$ &  $2$ \\ \hline
		$2_{001}$ &		$\left(\begin{array}{cc} -i&0 \\ 0&i \end{array}\right)$	& $0$ & $0$ & $0$ & $0$  \\ \hline
		$2_{010}$ &		$\left(\begin{array}{cc} 0&-1 \\ 1&0 \end{array}\right)$	& $0$ & $0$ & $0$ & $0$  \\ \hline
		$2_{100}$ &		$\left(\begin{array}{cc} 0&-i \\ -i&0 \end{array}\right)$	& $0$ & $0$ & $0$ & $0$  \\ \hline
		$4^{+}_{010}$ &		$\left(\begin{array}{cc} \sqrt{2}/2&-\sqrt{2}/2 \\ \sqrt{2}/2&\sqrt{2}/2 \end{array}\right)$	& $\sqrt{2}$ & $-\sqrt{2}$ & $\sqrt{2}$ & $-\sqrt{2}$ \\ \hline
		$2_{101}$ &		$\left(\begin{array}{cc} -i\sqrt{2}/2&-i\sqrt{2}/2 \\ -i\sqrt{2}/2&i\sqrt{2}/2 \end{array}\right)$	& $0$ & $0$ & $0$ & $0$  \\ \hline
		$4^{-}_{010}$ &		$\left(\begin{array}{cc} \sqrt{2}/2&\sqrt{2}/2 \\ -\sqrt{2}/2&\sqrt{2}/2 \end{array}\right)$	& $\sqrt{2}$ & $-\sqrt{2}$ & $\sqrt{2}$ & $-\sqrt{2}$ \\
\hline

	\end{tabular}
\end{table}

\subsection{Irreducible representations of $PbPt_3$}\label{app:b-pbpt3}

\restylefloat{table}
\begin{table}[H]
	\begin{tabular}{|c|c|}\hline
		$\nu$ & 68 \\ \hline \hline
		$n_{\Gamma}^{6}$ & $4$\\ \hline
		$n_{\Gamma}^{7}$ & $4$\\ \hline
		$n_{\Gamma}^{8}$ & $3$\\ \hline
		$n_{\Gamma}^{9}$ & $1$\\ \hline
		$n_{\Gamma}^{{10}}$ & $7$\\ \hline
		$n_{\Gamma}^{{11}}$ & $4$\\ \hline \hline
		$n_{M}^{6}$ & $10$\\ \hline
		$n_{M}^{7}$ & $9$\\ \hline
		$n_{M}^{8}$ & $8$\\ \hline
		$n_{M}^{9}$ & $7$\\ \hline \hline
		$n_{R}^{6}$ & $3$\\ \hline
		$n_{R}^{7}$ & $4$\\ \hline
		$n_{R}^{8}$ & $2$\\ \hline
		$n_{R}^{9}$ & $3$\\ \hline
		$n_{R}^{{10}}$ & $7$\\ \hline 
		$n_{R}^{{11}}$ & $4$\\ \hline  \hline
		$n_{X}^{6}$ & $9$\\ \hline
		$n_{X}^{7}$ & $9$\\ \hline
		$n_{X}^{8}$ & $9$\\ \hline
		$n_{X}^{9}$ & $7$\\ \hline
	\end{tabular}
\end{table}

\section{Double group}
\label{appPR}
Here we provide an example of projective representations to clarify the relation to the double group.
\subsection{Definition of the $\mathbb{Z}_2\times\mathbb{Z}_2$ group}
We discuss the point group $G=222$, which is isomorphic to $\mathbb{Z}_2\times\mathbb{Z}_2$. 
As a set, $G$ contains four elements: $G=\{e,X,Y,Z\}$, where $e$ is the identity and $X$, $Y$, and $Z$ are, respectively, the two-fold rotation about $x$, $y$, and $z$ axis.
The group product is defined by
\begin{eqnarray}
&X\cdot X=Y\cdot Y=Z\cdot Z=e,\\
&X\cdot Y=Y\cdot X=Z,\\
&Y\cdot Z=Z\cdot Y=X,\\
&Z\cdot X=X\cdot Z=Y.
\end{eqnarray}

\subsection{Linear representations of the $\mathbb{Z}_2\times\mathbb{Z}_2$ group}
Linear representations of $G$ must satisfy
\begin{eqnarray}
U(g)U(g')=U(gg')\quad \text{for } g,g'\in G.
\end{eqnarray}
There are four linear representations of $G$, which are all 1D representations. They are given by $U(e)=1$, $U(X)=\xi$, $U(Y)=\xi'$, and $U(Z)=\xi\xi'$, where $\xi=\pm1$ and $\xi'=\pm1$.

\subsection{Projective representation of the $\mathbb{Z}_2\times\mathbb{Z}_2$ group}
\label{appc3}
Projective representations of $G$ must satisfy
\begin{eqnarray}
U(g)U(g')=\omega(g,g')U(gg')\quad \text{for } g,g'\in G
\end{eqnarray}
for a projective factor $\omega$ obeying the co-cycle condition $\omega(g,g')\omega(gg',g'')=\omega(g,g'g'')\omega(g',g'')$ for $g,g',g''\in G$.

As an example, let us set
\begin{eqnarray}
&\omega(e,X)=\omega(e,Y)=\omega(e,Z)=+1,\\
&\omega(X,e)=\omega(Y,e)=\omega(Z,e)=+1,\\
&\omega(X,X)=\omega(Y,Y)=\omega(Z,Z)=-1,\\
&\omega(X,Y)=\omega(Y,Z)=\omega(Z,X)=-1,\\
&\omega(Y,X)=\omega(Z,Y)=\omega(X,Z)=+1.
\end{eqnarray}
There is only one projective representation for this particular $\omega$, which reads $U(e)=\sigma_0$, $U(X)=i\sigma_x$, $U(Y)=i\sigma_y$, $U(Z)=i\sigma_z$. Here, $\sigma_0$ is the 2D identity matrix and $\sigma_x$, $\sigma_y$, and $\sigma_z$ are the Pauli matrices.

The choice of $\omega$ is not unique. For example, one can instead use
\begin{eqnarray}
&\omega(X,Y)=\omega(Y,Z)=\omega(Z,X)=+1,\\
&\omega(Y,X)=\omega(Z,Y)=\omega(X,Z)=-1.
\end{eqnarray}
When all other components are unchanged, the corresponding projective representation is given by $U(e)=\sigma_0$, $U(X)=-i\sigma_x$, $U(Y)=-i\sigma_y$, $U(Z)=-i\sigma_z$. Other possible choices of $\omega$ can also be generated by the replacement $U(g)\rightarrow e^{i\theta(g)}U(g)$ and $\omega(g,g')\rightarrow e^{i[\theta(g)+\theta(g')-\theta(gg')]}\omega(g,g')$.  ($\theta(g)$ must be chosen in such a way that the co-cycle condition is respected.) 

\subsection{Definition of the double group of the $\mathbb{Z}_2\times\mathbb{Z}_2$ group}
In the double group approach, we consider the linear representation of an enlarged group $G'$. $G'$ contains both $+g$ and $-g$ for each element of $G$. Hence, $G'=\{\eta g|\eta=\pm1,g\in G\}$ has eight elements in total. Given a projective factor $\omega$, the group product of $G'$ is defined by
\begin{eqnarray}
(\eta g)\cdot (\eta' g')=(\eta\eta'\omega(g,g')g\cdot g')\in G' \quad \text{for } g,g'\in G \text{ and }\eta,\eta'=\pm1.
\end{eqnarray}
Clearly, the very definition of the double group (i.e., the product rule itself) depends on the choice of $\omega$.

\subsection{Linear representation of the double group of the $\mathbb{Z}_2\times\mathbb{Z}_2$ group}
Let us assume our first choice of $\omega$ above.  There are four 1D linear representations of $G'$, given by $U'(\eta e)=1$, $U'(\eta X)=\xi$, $U'(\eta Y)=\xi'$, and $U'(\eta Z)=\xi\xi'$, where $\eta=\pm1$, $\xi=\pm1$, and $\xi'=\pm1$. There is also one 2D linear representation of $G'$, given by $U'(\eta e)=\eta\sigma_0$, $U'(\eta X)=\eta i\sigma_x$, $U'(\eta Y)=\eta i\sigma_y$, $U'(\eta Z)=\eta i\sigma_z$ for $\eta=\pm1$. 

In the double group treatment of the projective representation, one must choose linear representations of $G'$ that satisfy $U'(\eta g)=\eta U'(+g)$ for every $g\in G$ and $\eta=\pm1$. Hence in this example, the valid choice is given by the 2D representation. The projective representation of $G$ we obtained in~\ref{appc3} can be reproduced by setting $U(g)=U'(+g)$ for each $g\in G$.

\bibliography{ref}

\end{document}